\documentclass[journal]{IEEEtran} 

\usepackage[cmex10]{amsmath}
\usepackage{amssymb}
\usepackage{color}
\usepackage{enumerate}
\usepackage{wasysym}

\DeclareMathOperator*{\argmax}{arg\,max}

\let\abs=\envert

\newcommand{\mbf}[1]{\mathbf{#1}}

\usepackage{graphicx}
\usepackage{hyperref}
\usepackage{threeparttable}
\usepackage{array,booktabs,tabularx}
\usepackage{multirow}

\usepackage{algorithmic}

\usepackage{array}

\usepackage{color}
\usepackage{tabularx}
\usepackage[switch,modulo]{lineno}

\usepackage[tight,footnotesize]{subfigure}

\usepackage{epsfig}
\usepackage{epstopdf}
\usepackage{xr}
\begin{document}

\title{Estimating Uncertainty in Neural Networks for Cardiac MRI Segmentation: 
A Benchmark Study}

\author{Matthew Ng, Fumin Guo, Labonny Biswas, Steffen E. Petersen, Stefan K. Piechnik, Stefan Neubauer, Graham Wright
\thanks{Corresponding authors: Fumin Guo (fguo@hust.edu.cn), Matthew Ng (matthewng.ng@mail.utoronto.ca)}
\thanks{M Ng and G Wright are with the Physical Sciences Platform
at Sunnybrook Research Institute and Department of Medical Biophysics,
University of Toronto, Toronto, ON, Canada.}
\thanks{F Guo is with the Wuhan National Laboratory for Optoelectronics 
and Biomedical Engineering, Huazhong University of Science and 
Technology, Wuhan, China, and Physical Sciences Platform
at Sunnybrook Research Institute and Department of Medical Biophysics,
University of Toronto, Toronto, ON, Canada.}
\thanks{L Biswas is with the Physical Sciences Platform, Sunnybrook Research
Institute, Toronto, ON, Canada.}
\thanks{SE Petersen is with the William Harvey Research Institute, NIHR Barts
Biomedical Research Centre, Queen Mary University of London, UK and Barts Heart
Centre, St Bartholomew’s Hospital, Barts Health NHS Trust, West Smithfield,
London, UK.} 
\thanks{SK Piechnik and S Neubauer are with the Division of Cardiovascular
Medicine, Oxford NIHR Biomedical Research Centre, Radcliffe Department of
Medicine, University of Oxford, Oxford, UK.}
\thanks{Copyright \copyright 2022 IEEE. Personal use of this material is 
permitted. Permission from IEEE must be obtained for all other uses, 
in any current or future media, including reprinting/republishing this 
material for advertising or promotional purposes, creating new collective 
works, for resale or redistribution to servers or lists, or reuse of any 
copyrighted component of this work in other works.
DOI: 10.1109/TBME.2022.3232730}
}

\markboth{IEEE TRANSACTIONS ON BIOMEDICAL ENGINEERING, VOL. XX, NO. XX, 202X}
{Ng \MakeLowercase{\textit{et al.}}: Estimating Uncertainty in Neural Networks for Cardiac MRI Segmentation: 
A Benchmark Study}

\maketitle

\begin{abstract}
Objective: Convolutional neural networks (CNNs) have demonstrated promise in automated
cardiac magnetic resonance image segmentation. However, when using CNNs in a
large real-world dataset, it is important to quantify segmentation uncertainty
and identify segmentations which could be problematic. In this work, we
performed a systematic study of Bayesian and non-Bayesian methods for estimating
uncertainty in segmentation neural networks. 

Methods: We evaluated Bayes by Backprop,
Monte Carlo Dropout, Deep Ensembles, and Stochastic Segmentation
Networks in terms of segmentation accuracy, probability calibration,
uncertainty on out-of-distribution images, and segmentation quality control. 

Results: We observed that
Deep Ensembles outperformed the other methods except for images with heavy noise
and blurring distortions. We showed that Bayes by Backprop 
is more robust to noise distortions while Stochastic Segmentation Networks 
are more resistant to blurring distortions. For segmentation quality control, we
showed that segmentation uncertainty is correlated with segmentation accuracy
for all the methods. With the incorporation of uncertainty estimates, we were
able to reduce the percentage of poor segmentation to 5\% by flagging 31--48\% of
the most uncertain segmentations for manual review, substantially lower than
random review without using neural network uncertainty (reviewing 75--78\% of all images).

Conclusion: This work provides a comprehensive evaluation of uncertainty 
estimation methods and showed that Deep Ensembles outperformed other methods
in most cases.

Significance: Neural network uncertainty measures can help identify 
potentially inaccurate segmentations and alert users for manual review.
\end{abstract}


\IEEEpeerreviewmaketitle

\section{Introduction}
\IEEEPARstart{C}{ardiac} magnetic resonance imaging (MRI) is the gold
standard for evaluating cardiac function due to its excellent soft tissue
contrast, high spatial and temporal resolution, and non-ionizing radiation
\cite{Peng2016}. Segmentation of cardiac structures such as the left ventricle
cavity, left ventricle myocardium, and right ventricle cavity is required as a
first step to quantify clinically relevant imaging biomarkers, such as the left
ventricle ejection fraction and myocardial mass. Recently, convolutional neural
networks (CNNs) have demonstrated promise for automatic cardiac MR image
segmentation \cite{bai2018automated, bernard2018deep}
and may facilitate the development of efficient cardiac MR image processing 
pipelines for research and clinical use.
However, when using a CNN in an automated image analysis pipeline, it is
important to automatically identify which segmentations are problematic and
require further manual inspection. This may improve workflow efficiency by
focusing only on problematic cases, avoiding the review of all images and
reducing errors in downstream analysis.

This problem has been referred to as \textit{segmentation quality control} and
is closely related to the task of anomaly detection or out-of-distribution
detection \cite{bishop1994novelty}. While there are several 
approaches to this problem (e.g., using a dedicated quality control module \cite{galati2022accuracy}), 
in this work, we focus on the approach of using 
predictive uncertainty estimates of a segmentation model to solve this problem.
The main idea here
is that segmentation outputs with low uncertainty are \textit{likely} correct
while those with high uncertainty are \textit{likely} problematic. While
several studies have attempted to estimate CNN segmentation uncertainty, most of
them used Monte Carlo (MC) Dropout or Deep Ensembles. However, there are some
limitations associated with these methods, which motivates us to explore other
algorithms. For example, when using a fixed dropout rate in MC Dropout, the
model uncertainty does not decrease when training data is increased. This is
potentially problematic since model uncertainty should approach zero in the
limit of infinite training data \cite{osband2016risk}. 
For Deep Ensembles, it is not clear \emph{why}
this method generates well-calibrated uncertainty estimates.
While Deep Ensembles was shown to learn diverse functions \cite{fort2019deep}, recent work \cite{abe2022deep}
has shown that ensemble diversity does not explain the improved uncertainty 
estimates on out-of-distribution data. In addition, in
previous studies, evaluation of these algorithms was mostly limited to
correlations between the predictive uncertainty and segmentation accuracy or
metrics measuring how uncertainty can be used to improve segmentation
\cite{roy2018inherent,jungo2018uncertainty,sander2019towards}.
There is a need to have a proper benchmark and metrics to 
evaluate uncertainty from different methods.
In this work, we performed a systematic evaluation of several
Bayesian and non-Bayesian approaches for uncertainty estimation. In particular,
we evaluated Bayes by Backprop, MC Dropout, Deep Ensembles, and
Stochastic Segmentation Networks based on segmentation accuracy, probability
calibration, uncertainty on out-of-distribution datasets, and finally, we
demonstrated the utility of these methods for segmentation quality control.

\subsection{Uncertainty in Neural Networks}
Uncertainty is usually classified into epistemic uncertainty or 
aleatoric uncertainty \cite{kendall2017uncertainties}.
Epistemic or model uncertainty is uncertainty in model parameters due 
to a finite amount of data. In contrast, aleatoric or data-dependent uncertainty
is uncertainty due to the data itself and cannot be reduced even with more data.
This distinction can be useful when thinking about sources of uncertainty;
however, it is difficult to distinguish these two types in practice. 
Instead, we can think about modelling uncertainties in neural networks by learning a
distribution of neural network weights or by learning a distribution of neural
network outputs for each individual input \cite{kendall2017uncertainties}.
Uncertainty from both types of models can be used and compared for downstream tasks.

Bayesian neural networks (BNNs) provide a theoretical framework for
generating well-calibrated uncertainty estimates \cite{neal2012bayesian}. In
BNNs, we are interested in learning the posterior distribution of the neural 
network weights instead of a maximum likelihood or maximum-a-posteriori estimate. A
challenge in learning BNNs is that integration over the posterior is intractable
in high dimensional space. As such, inference techniques such as stochastic
variational inference are commonly used as approximation. Examples include
variational dropout \cite{kingma2015variational}, MC Dropout
\cite{gal2016dropout}, Bayes by Backprop \cite{blundell2015weight},
multiplicative normalizing flows \cite{louizos2017multiplicative}, and Flipout
\cite{wen2018flipout}. Non-Bayesian methods for estimating uncertainty in neural
networks include bootstrapping \cite{efron1986bootstrap}, Deep
Ensembles \cite{lakshminarayanan2017simple}, and Resampling Uncertainty Estimation
\cite{schulam2019can}. These methods estimate changes to the neural network 
when it is trained on different samples from the 
same training distribution. Note that in Bayesian methods, uncertainty is learned
during training and is tightly coupled to the model structure. In non-Bayesian
methods, uncertainty is learned during training or estimated after training.
Another approach to uncertainty estimation is to directly learn a distribution of
neural network outputs (and/or intermediate feature maps) instead of the
weights. This is usually achieved by parameterizing the output distribution and
learning the parameters during training as shown in \cite{kendall2017uncertainties}.

\subsection{Related Studies}

The majority of investigations of BNNs used MC Dropout to approximate the
posterior distribution of the weights and exploration of ways to evaluate the
quality of uncertainty has been limited. Previous studies
\cite{roy2018inherent}, \cite{jungo2018uncertainty}, and
\cite{wang2019automatic} used MC Dropout for brain structure, brain tumour, and
brain tumour cavity segmentation. These studies reported positive correlations
between segmentation accuracy and uncertainty measures. Nair et al.
\cite{nair2019exploring} compared different uncertainty measures in brain lesion
segmentation and showed that uncertainty measures can be used to improve lesion
detection accuracy. Sander et al. \cite{sander2019towards} applied MC Dropout
for cardiac MRI segmentation and showed that training a CNN using a Brier loss
or cross-entropy loss produced well-calibrated pixel-wise segmentation
uncertainty, and correcting uncertain pixels can improve segmentation results
consistently. Devries et al. \cite{devries2018leveraging} used MC Dropout and
non-Bayesian methods to generate skin lesion segmentation and segmentation
uncertainty maps, which were then entered into another neural network to predict
the Jaccard index of the segmentation. Hann et al. \cite{hann2019quality}
estimated the quality of aortic MRI segmentation using an ensemble of neural
networks and demonstrated improved segmentation accuracy with the use of these
segmentation quality estimates. More recently, Jungo et al.
\cite{jungo2019assessing} compared MC Dropout, Deep Ensembles, and auxiliary
networks for predicting pixel-wise segmentation errors for two medical image
segmentation tasks. In addition to segmentation probability calibration, they
examined the overlap between segmentation uncertainty and errors, and the
fraction of images which would benefit from uncertainty-guided segmentation
correction. In a follow-up study \cite{jungo2020analyzing}, the authors compared
different aggregation methods for uncertainty measures and their performance for
segmentation failure detection. Similarly, Mehrtash et al.
\cite{mehrtash2020confidence} compared MC Dropout and Deep Ensembles for CNNs
trained with different loss functions in terms of probability calibration and
correlation between segmentation accuracy and uncertainty measures. However,
these studies did not evaluate other Bayesian methods such as Bayes by Backprop
and the performance of these methods on out-of-distribution datasets is unknown.
Other methods such as Probabilistic U-net
\cite{kohl2018probabilistic}, PHiSeg \cite{baumgartner2019phiseg}, and
Stochastic Segmentation Networks \cite{monteiro2020stochastic} estimate 
uncertainty by directly predicting a distribution of neural network outputs.
These methods have been applied to brain and lung tumour segmentation and have
shown to produce diverse outputs matching inter-observer manual segmentation variability.
\cite{puyol2020automated} presented a framework for segmentation quality 
control of cardiac T1 maps using the evidence lower bound scores from PHiSeg and a separate 
quality control neural network. While this showed great sensitivity and specificity
for detecting poor segmentations, 
it is not clear how other algorithms would compare with PHiSeg for this task.

\subsection{Contributions}
In this
work, we performed a systematic study of Bayesian and non-Bayesian neural
networks for estimating uncertainty in the context of cardiac MRI segmentation.
Our contributions are summarized as follows:

\begin{enumerate}
\item We compared MC Dropout and Deep Ensembles with Bayes by
Backprop, which is a more theoretically justified algorithm for learning
uncertainty in BNNs. We performed a comprehensive evaluation of these algorithms
in terms of segmentation accuracy, probability calibration, uncertainty on
out-of-distribution datasets, and segmentation quality control.

\item  We evaluated MC Dropout, Deep Ensembles, Bayes by Backprop, and
Stochastic Segmentation Networks on cardiac MRI datasets with
various degrees of noise, blurring, and stretching distortions to mimic complex
clinical scenarios and to investigate the relationships between image
distortions and neural network uncertainty estimates. We showed that Bayes by
Backprop is more robust to noise distortions while Stochastic
Segmentation Networks are more resistant to blurring distortions.

\item We introduced a novel area-under-the-curve metric for quantifying
algorithm performance on segmentation quality control. We showed that with the
use of Deep Ensemble uncertainty estimates, 31--48\% of the most uncertain
segmentations need to be reviewed to reduce the percentage of poor segmentations
to 5\%, whereas $\sim$80\% of the results need to be reviewed when neural
network uncertainty measures are not used.

\end{enumerate}

We hope this work will serve as a benchmark for evaluating uncertainty in
cardiac MRI segmentation and inspire further work on uncertainty estimation in
medical image segmentation.
\graphicspath{ {./plots/} }

\section{Methods}

\subsection{Bayesian Neural Networks} \label{sec:bnn-intro}
Given a dataset of $N$ images $\mathbf{X}= \{\mathbf{x}_i\}$, $i \in [1, N]$, 
and the corresponding manual segmentation $\mathbf{Y}= \{\mathbf{y}_{i}\}$ with
$C$ classes, we fit a neural network parameterized by weights $ \mathbf{w} $ to
perform segmentation. In BNNs, we are interested in learning the posterior
distribution of the weights $ p(\mathbf{w} | \mathbf{X}, \mathbf{Y}) $, instead
of a maximum likelihood or maximum-a-posteriori estimate. This posterior
distribution represents uncertainty in the weights, which could be propagated to
calculate uncertainty in the predictions \cite{mackay1995probable}. In addition,
BNNs have been shown to be able to improve the generalizability of neural
networks \cite{mackay1995probable}.

A challenge in learning BNNs is that calculating the posterior is intractable
due to the high dimensionality of the weights. Variational inference
\cite{hoffman2013stochastic} is a scalable technique that aims to learn an
approximate posterior distribution of the weights $q(\mathbf{w})$ by minimizing
the Kullback-Leibler (KL) divergence between the approximate and true
posterior. This is equivalent to maximizing the evidence lower bound as follows:
\begin{equation}
\label{eq:elbo}
\argmax_{q(\mathbf{w})} {\mathbb{E}_{q(\mathbf{w})}[\log
p(\mathbf{Y}| \mathbf{X}, \mathbf{w})] - \lambda \cdot
\textrm{KL}[q(\mathbf{w}) || p(\mathbf{w})]} \,,
\end{equation}
where $\mathbb{E}_{q(\mathbf{w})}[\cdot]$ denotes expectation over the
approximate posterior $q(\mathbf{w})$, $\log p(\mathbf{Y}| \mathbf{X},
\mathbf{w})$ is the log-likelihood of the training data with given weights
$\mathbf{w}$, $p(\mathbf{w})$ represents the prior distribution of $\mathbf{w}$,
and KL[$\cdot$] is the KL divergence between two probability distributions
weighted by a hyperparameter $\lambda >0$.

State-of-the-art segmentation neural networks such as the U-net formulate image
segmentation as a pixel classification problem. For each pixel $x_{i,j}$ in
image $\mathbf{x}_i$, $i \in [1, N]$, $j\in \Omega$,  the neural network
generates a prediction $\hat{y}_{i,j}$ with probability $p(\hat{y}_{i,j}=c), c \in [0, C-1]$, through softmax activation of the features in the last
layer. Assuming pixels are independent from each other, the
log-likelihood of the training data in Eq. \eqref{eq:elbo} is given by:
\begin{equation*}
\log p(\mathbf{Y} | \mathbf{X}, \mathbf{w}) = 
\sum_{i=1}^{N} \sum_{j \in \Omega} 
\sum_{c=0}^{C-1} [y_{i,j} = c] \cdot \log p(\hat{y}_{i,j} = c)\ ,
\end{equation*}
where $y_{i,j}$ is the manual label for pixel $j$ in image $x_i$
and [] is the indicator function. In this setting, the log-likelihood is also
the negative cross entropy between the manual segmentation and algorithm 
prediction. The prediction $\mathbf{\hat{y}}$ of a test image $\mathbf{x}$
is generated by marginalizing out the weights of the neural network, i.e.,
\begin{equation}
\label{eq:pred-marginalization}
p(\mathbf{\hat{y}} | \mathbf{x}) = 
\mathbb{E}_{q(\mathbf{w})}[p(\mathbf{\hat{y}}|\mathbf{x}, \mathbf{w})]\ ,
\end{equation}
where $p(\mathbf{\hat{y}}|\mathbf{x}, \mathbf{w})$
denotes the prediction of an image $\mathbf{x}$ given network weights
$\mathbf{w}$. In the following sections, we introduce methods for estimating an
approximate posterior $q(\mathbf{w})$ for the weights of a BNN.

\subsubsection{Bayes by Backprop}
One way to parameterize the approximate posterior $q(\mathbf{w})$ is to use a 
fully factorized Gaussian. 
In a fully factorized Gaussian, each weight $w$ in $\mathbf{w} $ is
independent from others and follows its own Gaussian distribution with mean
$\mu$ and standard deviation $\sigma$. To ensure $\sigma > 0$ and training
stability, $\sigma$ is parameterized by a real number $\rho$, i.e., $\sigma =
\textrm{softplus}(\rho) = \ln (1+e^\rho)$. The prior distribution
$p(\mathbf{w})$  is usually chosen as a fully factorized Gaussian with mean
$\mu_{prior}\mathbf{I}$ and covariance $\sigma_{prior}\mathbf{I}$, i.e.,
$p(\mathbf{w}) = \mathcal{N}(\mu_{prior}\mathbf{I}, \sigma_{prior}\mathbf{I})$,
where $\mathbf{I}$ represents an identity matrix. Gradient updates can be
performed using the ``reparameterization trick''. The training procedure is
known as Bayes by Backprop (BBB) \cite{blundell2015weight} and is briefly described
below:

\begin{enumerate}
\renewcommand{\labelenumi}{(\theenumi)}
\item For each weight $w$, sample $\epsilon \sim \mathcal{N}(0,1)$ and set $w =
\mu + \textrm{softplus}(\rho) \cdot \epsilon$. 
\item Calculate the loss based on Eq. \eqref{eq:elbo}, i.e.,
$-\log p(\mathbf{Y} | \mathbf{X}, \mathbf{w})
+ \lambda \cdot \textrm{KL}[q(\mathbf{w}) || \mathcal{N}(\mu_{prior}\mathbf{I},
\sigma_{prior}\mathbf{I})]$ .
\item Update $\mu$ and $\rho$ through gradient descent.

\end{enumerate}

After training, each weight $w$ can be sampled from $\mathcal{N}(\mu, \sigma)$,
which is then used to generate the segmentation predictions following Eq.
\eqref{eq:pred-marginalization}.

\subsubsection{MC Dropout}
MC Dropout (MCD) \cite{gal2016dropout} is another commonly used method for
learning BNNs because it is straightforward to implement and does not require
additional parameters or weights. MCD can be interpreted as choosing the
approximate posterior distribution $q(\mathbf{w})$ to be a mixture of two
Gaussians with minimal variances, e.g., one at $0$ and the other at the weight
$w$. Dropout is applied during training and testing to sample weights from $
q(\mathbf{w})$. In this method, the dropout rate is a hyperparameter chosen
empirically based on a validation dataset. The dropout rate defines the amount
of uncertainty in the weights and is fixed throughout network training and
testing.

\subsection{Deep Ensembles}
In addition to BNNs, we characterized and evaluated uncertainty estimates using
an ensemble of neural networks, i.e., Deep Ensembles
\cite{lakshminarayanan2017simple}. Deep Ensembles consist of multiple neural
networks trained using the same data (or different subsets of the same data)
with different random initializations. Combining these models in an ensemble has
been shown to produce well-calibrated probabilities in computer vision tasks and
the variability between the model predictions can be used to calculate
predictive uncertainty. This non-Bayesian method was inspired by the idea of
bootstrapping, where stochasticity in the sampling of the training data and
training algorithm define model uncertainty. This approach differs from Bayesian
methods since it does not require approximation of the posterior distribution of
the weights.

\subsection{Stochastic Segmentation Networks}
Another method for uncertainty estimation involves
predicting a distribution over the neural network logits before transforming
them into probabilities. Since standard segmentation neural networks output
pixelwise logits, a simple method is to assume a Gaussian distribution 
of the logits and that each pixel is independent from others, i.e., 
for each image $\mathbf{x}$, the neural network predicts logits 
$\mathbf{\eta} \sim \mathcal{N}(\mu(\mathbf{x}), \Sigma(\mathbf{x}))$, 
where $\mu(\mathbf{x}) \in \mathbb{R}^{\abs{\Omega}C}$ is the mean logit and 
$\Sigma(\mathbf{x}) \in \mathbb{R}^{\abs{\Omega}C \times \abs{\Omega}C}$ is a
diagonal covariance matrix \cite{kendall2017uncertainties}. Stochastic
Segmentation Networks (SSNs) \cite{monteiro2020stochastic} improved this method
by using a low rank multivariate normal distribution on the pixelwise logits to
model the dependencies between pixels in an image. In particular, SSNs use $
\Sigma(\mathbf{x}) = PP^T + D$, where $P \in \mathbb{R}^{\abs{\Omega}C\times R}$ is a low
rank matrix and $D \in \mathbb{R}^{\abs{\Omega}C \times \abs{\Omega}C} $ is a
diagonal matrix.

\subsection{Algorithm Evaluation}

We evaluated the uncertainty estimation algorithms based on three aspects: (1)
segmentation accuracy and probability calibration; (2) uncertainty on
out-of-distribution datasets; and (3) application of uncertainty estimates for
segmentation quality control. The purposes of these evaluations are as follows:

\begin{enumerate}
\renewcommand{\labelenumi}{(\theenumi)}
\item We show that BNNs can provide segmentation accuracies that are similar to or
higher than plain or point estimate neural networks. In addition, predicted
segmentation probabilities should be well-calibrated, i.e., a pixel with
predicted probability of 60\% belonging to the myocardium is 60\% myocardium according to some ground truth. From a
frequentist perspective, this means that out of all predictions with 60\%
probability, 60\% of the predictions are correct.
\item We measure segmentation uncertainty on out-of-distribution data to
validate that the uncertainty measures perform as expected, i.e., uncertainty
should increase when test datasets substantially differ from training
datasets.
\item We expect uncertainty measures to be useful in identifying problematic 
segmentations that require manual editing.
\end{enumerate}

We used the following metrics for these evaluations:

\subsubsection{Segmentation Accuracy}
We calculated the algorithm segmentation accuracy using Dice similarity
coefficient, average symmetric surface distance (ASSD), and Hausdorff distance
(HD), as previously described \cite{bai2018automated}.

\subsubsection{Probability Calibration}
These metrics measure how closely the neural network segmentation probabilities
match the manual segmentation probabilities on a per-pixel basis.

Following the notation in Sec. \ref{sec:bnn-intro}, let $\hat{y}_j$ and
$y_j$ denote the prediction and manual label of pixel $j$ in a given
image, $j\in \Omega$, respectively. We use $p(\hat{y}_j = c)$ 
to denote the average per-pixel probability from the samples of the neural network,
i.e., $p(\hat{y}_j = c) = \mathbb{E}_{q(\mathbf{w})}[p(\hat{y}_j = c | \mathbf{w})]$.
{Negative log-likelihood (NLL)}
measures how well the learned model fits the observed (testing) data and is
calculated as follows:
\begin{align*}
	\textrm{NLL} = \sum_{j\in \Omega} \sum_{c=0}^{C-1} [y_j=c] \cdot \log p(\hat{y}_j = c)\,.
\end{align*}
Note that NLL is sensitive to tail probabilities; that is, a model that
generates low probability for the correct class is heavily penalized.

Brier score (BS) \cite{brier1950verification} is a proper scoring rule used to measure probability calibration. It measures the mean squared error between the predicted
and manual segmentation probabilities:
\begin{align*}
\textrm{BS} = \sum_{j \in \Omega} \sum_{c=0}^{C-1} [p(\hat{y}_j = c) - p(y_j = c)]^2\,.
\end{align*} 
A Brier score of 0 indicates that the model is perfectly calibrated.

\subsubsection{Predictive Uncertainty Measures}
Predictive uncertainty can be calculated from neural network predictions to
indicate the degree of uncertainty of the outputs. This can be calculated per
pixel or per structure/class.

\paragraph{Pixelwise Uncertainty Measures}
Pixelwise uncertainty measures are calculated per pixel and averaged across all
pixels in an image if an image-level measure is required. In this work, we used
multi-class predictive entropy and multi-class mutual 
information as suggested to be superior in \cite{camarasa2021quantitative}:

\begin{itemize}
  \item {Multi-class Predictive Entropy} measures the spread of probabilities across
  all the classes in the mean prediction, i.e., 
	\begin{align*}
	\sum_{j \in \Omega} 
			  \sum_{c=0}^{C-1} \left[-p \left( \hat{y}_j = c \right) \log p \left(
			  \hat{y}_j = c \right) \right]\,.
	\end{align*}
  \item {Multi-class Mutual Information (MI)} measures how different each sample is from the 
  mean prediction and is calculated as:  
	\begin{align*}
	\mathbb{E}_{q(\mbf{w})} & \left [ \sum_{j\in \Omega} \sum_{c=0}^{C-1}
	\, p \left ( \hat{y}_j = c | \mbf{w} \right ) \, \log p \left ( \hat{y}_j =
	c|\mbf{w} \right) - \right .\\
	& \left . \sum_{j\in \Omega} \sum_{c=0}^{C-1} \, 
	p \left( \hat{y}_j = c \right) \, \log p \left( \hat{y}_j = c \right) \, \right ] \,,
	\end{align*}
	where $p \left( \hat{y}_j = c | \mbf{w} \right)$ denotes the prediction given a
	set of weights $\mathbf{w}$. MI is high if there are samples with both high
	and low confidence, and is low if all samples have low confidence or high
	confidence.

\end{itemize}

\paragraph{Structural Uncertainty Measures}
We define two structural uncertainty measures, which quantify how different the
prediction samples are for each structure in terms of Dice and ASSD. 
\begin{itemize}
	\item $\textrm{Dice}_\textrm{WithinSamples} \ (\textbf{Dice}_\textbf{WS})
	=\frac{1}{T} \sum_{i=1}^{T} \textrm{Dice}(\bar{S}, S_{i})$

	\item $\textrm{ASSD}_\textrm{WithinSamples} \ (\textbf{ASSD}_\textbf{WS}) 
	=\frac{1}{T} \sum_{i=1}^{T} \textrm{ASSD}(\bar{S}, S_{i})$\,,
\end{itemize}
where $\bar{S}$ is the mean of the $T$ segmentation predictions $S_{i},  
i \in [1, T]$. We expect
structural uncertainty measures to better align with common segmentation
accuracy metrics because of their global image-level focus.

\subsection{Datasets}

\subsubsection{UK BioBank (UKBB)} 
The UKBB dataset \cite{petersen2015uk} consists of 4845 healthy volunteers. 
For each subject, 2D cine cardiac MR images were acquired on a 1.5T Siemens
scanner using a bSSFP sequence under breath-hold conditions with ECG-gating
(pixel size = 1.8-2.3 mm, slice thickness = 8 mm, number of slices = $\sim$10,
number of phases = $\sim$50). Manual segmentation of the left ventricle blood
pool (LV), left ventricle myocardium (Myo), and right ventricle (RV) was
performed on the end-diastolic (ED) and end-systolic (ES) phases by one of eight
observers followed by random checks by an expert to ensure segmentation quality
and consistency. The dataset was randomly split into 4173, 103, and 569 subjects
for training, validation, and testing, respectively. We have permission to 
use the UKBB dataset through UK Biobank's generic RTB 
approval from the NHS North West REC.

\subsubsection{Automated Cardiac Diagnosis Challenge (ACDC)}
The ACDC dataset \cite{bernard2018deep} consists of 100 patients with one of 
five conditions: normal, myocardial infarction, dilated cardiomyopathy,
hypertrophic cardiomyopathy, and abnormal right ventricle. 2D cine MR images
were acquired using a bSSFP sequence on a 1.5T/3T Siemens scanner (pixel size =
0.7-1.9 mm, slice thickness = 5-10 mm, number of slices = 6-18, number of phases
= 28-40). Manual segmentation was performed at ED and ES phases with approval by
two experts. This dataset was used for testing only.

\subsection{Training Details}
We used a plain 2D U-net \cite{ronneberger2015u} for BBB, MCD, Deep
Ensembles, and SSN. The plain U-net consisted of 10 layers with
$3\times3$ filters and 2 layers with $1\times1$ convolutions followed by a
softmax layer. The number of filters ranged from 32 to 512 from the top to the
bottom layers.

For BBB, we experimented with different standard deviations of the prior distributions:
$\sigma_{prior}=$ 0.1 or 1.0 and different weights for the KL term: $\lambda=$
0.1, 1.0, 10, 30. These are commonly used hyperparameters in the literature
\cite{kendall2017bayesian, blundell2015weight, fortunato2017bayesian}.
For MCD, we added dropout on all layers or only on the middle layers of the
U-net with different dropout rates: 0.5, 0.3, and 0.1. These settings
effectively tuned the amount of uncertainty in the model. 
For both BBB and MCD, the final prediction was obtained by averaging the softmax
probabilities of T=50 samples. For Deep Ensembles, we trained 10 plain U-net
models separately using all the training data with different random
initializations and averaged the softmax probabilities of the 10
models. For SSN, we used a rank of 10 for the multivariate 
normal distribution of the logits, as suggested in \cite{monteiro2020stochastic}.
More training details can be found in Supplementary Material
Section I.

\graphicspath{ {./plots/} }

\section{Experiments and Results}

\begin{table*}[!htp]
  \centering
  \begin{threeparttable}
  \caption{Segmentation accuracy and probability calibration of the plain U-net,
  U-net with MC Dropout, BBB, Deep Ensemble, or SSN on n=1138 images from
  UKBB. $\uparrow$ indicates higher is better. $\downarrow$ indicates lower is
  better. Format: Mean (Standard Deviation).}
  \vspace{-0.4cm}
  \label{tab:segmentation_performance}

  \setlength\tabcolsep{2pt}
  \fontsize{7.5pt}{8.5pt}\selectfont
  \begin{tabularx}{1.0\textwidth}{lccccccccccc}
  \toprule
    				   & \multicolumn{3}{c}{Dice $\uparrow$}& \multicolumn{3}{c}{ASSD (mm) $\downarrow$} & \multicolumn{3}{c}{HD (mm) $\downarrow$} & NLL  $\downarrow$  					& 	BS  $\downarrow$	\\
    \cmidrule(l{5pt}r{5pt}){2-4} \cmidrule(l{5pt}r{5pt}){5-7} \cmidrule(l{5pt}r{5pt}){8-10}
    			   	 &    LV 	    &   Myo   	&    RV     &    LV      &  Myo  	&   RV     &   LV      &  Myo  	&   RV    & $(\times 10^{-2})$					& 	$(\times 10^{-3})$			\\
    \midrule
     Plain	U-net &	$.941(.038)$  & 	$.882(.031)$	&  $.907(.043)$  &     $1.01(.38)$  &	$1.02(.43)$   & 	$1.61(.69)$  	&  $2.98(0.99)$  & $3.83(1.39)$	&	$6.53(2.67)$	&	$1.11(.38)$		& $1.67(.54)$ 	\\
     BBB		   &	  $.942(.037)^*$  & 	$.883(.030)^*$ 	&   $.908(.043)^*$  & 	$1.00(.36)^*$  &	$1.00(.30)^*$  & 	$1.60(.70)^*$  &  $2.96(0.96)^*$	&	$3.80(1.23)^\dagger$	&	$6.39(2.59)^*$	&	$1.10(.35)^*$		& $1.66(.53)^*$ 	\\ 
     MCD-0.1   &    $.941(.037)^*$  & 	$.882(.030)^\dagger$ 	& 	$.907(.043)^*$  & 	$1.00(.36)^*$  &	$1.01(.31)^\dagger$  & 	$1.61(.70)^\dagger$  &  $2.97(0.97)^\dagger$  &	$3.82(1.24)^\dagger$	&	$6.49(2.59)^\dagger$	& $1.10(.36)^*$ 	& $1.66(.53)^*$	  \\
     MCD-0.5   &    $.940(.038)^*$  & 	$.879(.030)^*$ 	& 	$.906(.043)^*$  & 	$1.03(.38)^*$  &	$1.04(.31)^*$  & 	$1.64(.69)^*$  &  $3.05(1.03)^*$  &	$3.95(1.29)^*$	&	$6.64(2.61)^*$	& $1.13(.35)^*$	  & $1.70(.53)^*$ 	\\
     Ensemble  &    $\mathbf{.943(.037)^*}$	&   $\mathbf{.885(.030)^*}$  &	  $\mathbf{.909(.043)^*}$  & 	$\mathbf{0.98(.36)^*}$	 &  $\mathbf{0.98(.29)^*}$  &	$\mathbf{1.57(.72)^*}$	 &  $\mathbf{2.90(0.93)^*}$  & $\mathbf{3.68(1.20)^*}$	& $\mathbf{6.26(2.56)^*}$  & $\mathbf{1.07(.35)^*}$   & $\mathbf{1.63(.52)^*}$   \\
     SSN       &    $.940(.037)^*$  &   $.882(.030)^\dagger$  &   $.903(.043)^*$  &   $1.03(.39)^*$	 &  $1.03(.32)^*$	 &  $1.65(.67)^*$	 &  $3.08(1.10)^*$	& $3.90(1.29)^*$	& $6.64(2.55)^*$  & $1.15(.33)^*$	  & $1.73(.50)^*$   \\
   \bottomrule
  \end{tabularx}
  \begin{tablenotes}\footnotesize
    \item[*] statistically different compared to the Plain U-net (Wilcoxon signed-rank test, p $<$ 0.05, $N=1138$ images).
    \item[$\dagger$] not statistically different compared to the Plain U-net (Wilcoxon signed-rank test, p $>$ 0.05, $N=1138$ images).
  \end{tablenotes}

  \end{threeparttable}
\end{table*}

To select the hyperparameters for each method, we chose 
the models with the lowest NLL on the validation dataset since NLL is directly
related to segmentation accuracy and probability calibration.
For BBB, $\lambda=10$ and $\sigma_{prior}=0.1$ achieved the best NLL on the
validation dataset. For MCD, adding dropout in the middle layers with a dropout
rate of 0.1 (MCD-0.1) performed the best. We also reported results for MC
Dropout with a dropout rate of 0.5 in the middle layers (MCD-0.5), which is
commonly used in the literature.

\subsection{Segmentation Accuracy and Probability Calibration} 
\label{sec:results-accuracy-calibration}

As shown in Table \ref{tab:segmentation_performance}, Deep Ensembles performed
the best in terms of segmentation accuracy and probability calibration. This was
followed by BBB and MCD-0.1, which were comparable to the plain U-net. MCD-0.5
performed slightly worse than the other models. The differences 
of these methods compared to the plain U-net were small but mostly statistically 
significant except for some metrics between plain U-net and MCD-0.1 (Table \ref{tab:segmentation_performance}).
These results indicate that
Bayesian approaches or Deep Ensembles can yield similar, if not better,
segmentation results compared to a plain U-net while providing uncertainty
estimates at the same time. The tradeoff is that Deep Ensembles 
and BBB use more memory and computation time compared to a plain U-net.

Examples of predicted segmentation from all 
methods are shown in Supplementary Figure S1.

\subsection{Uncertainty on Distorted Images}
In order to validate uncertainty measures as indicators of
``out-of-distribution'' datasets, we applied the trained models to carefully
generated test images with various magnitudes of distortions,
including:

\begin{itemize}
  \item adding Rician noise, as found in MR images
  \cite{gudbjartsson1995rician}, with magnitudes ranging from 0.05 to 0.10
  (on normalized images with intensities ranging from 0 to 1),
  \item Gaussian blurring with a standard deviation of 1--4 pixels,
  \item deforming or stretching around LV, Myo, and RV. 
\end{itemize}

Note that these distortions were not applied as part of data augmentation during
training and these images were not seen by the neural networks. As such, we
expected decreased segmentation accuracy and increased predictive uncertainty in
images with greater magnitudes of distortions. Figure
\ref{fig:distortion_entropy_images} shows examples of distorted images.

\subsubsection{Trends with Increasing Distortions}

\begin{figure*}[!htp]
  \centering
  \includegraphics[width=0.9\textwidth]{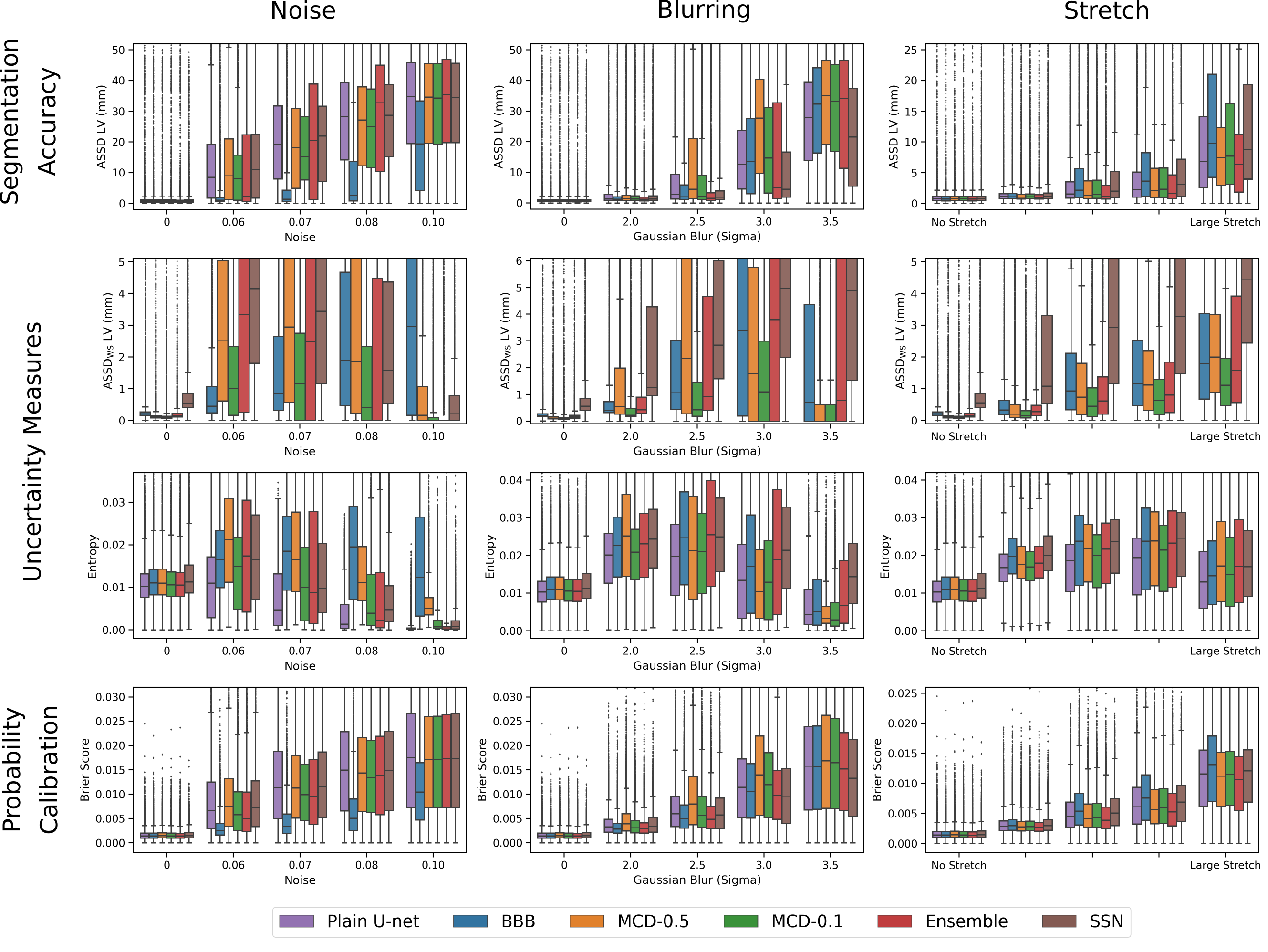}
  \caption{Segmentation accuracy (ASSD LV), uncertainty measures
  (ASSD$_\textrm{WS}$ LV, entropy), and probability
  calibration (Brier score) on images with increasing
  magnitude of noise, blurring, and stretching using a plain U-net, U-net with BBB, MC
  Dropout, Deep Ensemble, and SSN.}
  \label{fig:all_distortion_boxplots}
\end{figure*}

Figure~\ref{fig:all_distortion_boxplots} and Supplementary Table
S1 show that the segmentation accuracy decreased
as the magnitude of the distortions (noise, Gaussian blur, stretch) was
increased. This is expected since these types of distortions were not seen
during training and increasing the magnitude of the distortions results in
greater differences with the original training datasets.

\begin{figure*}[!ht]
  \centering
  \includegraphics[width=\textwidth]{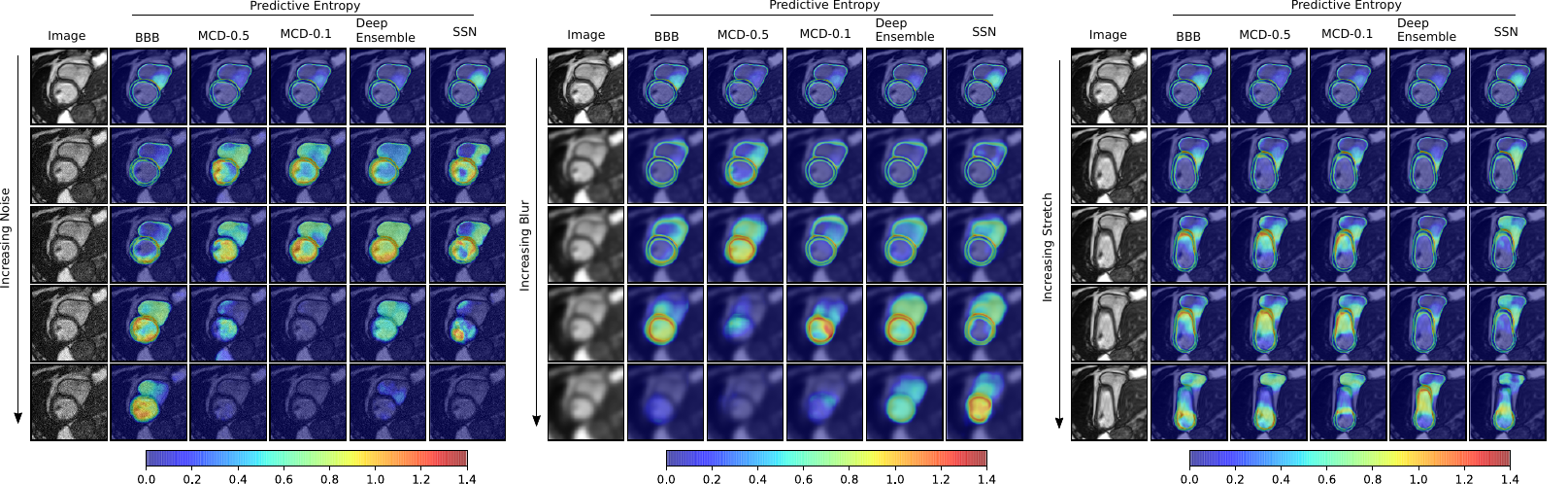}
  \caption{Segmentation predictive entropy on images with increasing noise,
  blurring, and stretching. BBB showed the highest uncertainty on images with
  heavy noise (last two rows) while SSN showed the highest uncertainty on images
  with heaving blurring (last two rows); Deep Ensembles showed slightly 
  higher uncertainty on images with heavy stretching compared to other 
  methods. Reproduced with permission of UK Biobank \copyright.}
  \label{fig:distortion_entropy_images}
\end{figure*}

In addition, we observed that the predictive uncertainty increased (i.e.,
higher predictive entropy, mutual information, ASSD$_\textrm{WS}$, and lower
Dice$_\textrm{WS}$) with increasing magnitude of distortions but this 
decreased after a certain threshold, as shown in
Figure~\ref{fig:all_distortion_boxplots}. This was the case for Deep Ensembles,
BBB, MCD, and SSN on images with noise and blurring distortions
but not with stretching. For example, for BBB, the median predictive entropy for
images with slight, moderate, and large additional noise was $1.66 \times
10^{-2}$, $1.95 \times 10^{-2}$, and $1.23 \times 10^{-2}$, respectively.
Similarly, the median ASSD$_\textrm{WS}$ was 0.40, 3.40, and 0.71 mm for images
with slight, moderate, and large amount of blurring, respectively (Supplementary
Table S3). Figure
\ref{fig:distortion_entropy_images} and Supplementary Figures
S2-S4 show
examples of segmentation predictions and uncertainty (predictive entropy, mutual
information) for all the methods on images with increasing noise, blurring, and
stretching.

While the increasing predictive uncertainty associated with increasing 
magnitude of distortions was expected, the decrease in predictive uncertainty
after a threshold in cases of noise and blurring distortions is surprising. 
Specifically, for images that were highly distorted, all pixels were classified
as background with low uncertainty (Figure~\ref{fig:distortion_entropy_images},
bottom row). Although this seems correct when only given the labeling choices of
background, LV, Myo, and RV, we argue that the distorted pixels are markedly
different from the background pixels in the training images and therefore,
should have high uncertainty nonetheless. This is a limitation of all the
uncertainty models tested and may be improved using more expressive posteriors.

Another observation is that the uncertainty measures began to fail/decrease
when dramatic segmentation errors occurred, as shown in Figure
\ref{fig:all_distortion_boxplots}. This suggests that other heuristics or
algorithms such as those presented in \cite{ruijsink2019fully} can be used to
complement the uncertainty measures when trying to detect inaccurate
segmentations. For example, segmentation with a non-circular LV blood pool or a
blood volume $<$ 50 mL is highly problematic and may indicate poor segmentation. 

\subsubsection{Comparison between Deep Ensembles, BBB, MC Dropout, and SSN}

In terms of segmentation accuracy and probability calibration, BBB was more
robust to noise distortions compared to the other methods. Specifically, BBB
showed higher Dice and lower ASSD for LV, lower NLL and BS on images with
greater noise distortions (Figure~\ref{fig:distortion_entropy_images}, rows 3-5
and Supplementary Tables S1 and
S2, degree of distortion = 2, 3, 4).
SSN showed higher segmentation accuracy in cases of greater
blurring while Deep Ensembles showed higher segmentation
accuracy with stretching distortions compared to the other methods
(Figure~\ref{fig:all_distortion_boxplots} and Supplementary Table
S1, degree of distortion = 3, 4).
In the greatest noise, blurring, and stretching distortions
tested, BBB, SSN, and Deep Ensembles had statistically higher
segmentation accuracy and probability calibration, respectively, when 
compared to other methods (Supplementary Tables S1 and
S2).

\subsection{Uncertainty on Dataset Shift}
To further validate the uncertainty measures, we applied the models trained on
the UKBB dataset to a distinctly different ACDC dataset. We expect decreased
segmentation accuracy and increased predictive uncertainty on the ACDC test
dataset compared to the UKBB test dataset, due to the presence of cardiac 
pathologies on the ACDC dataset and slightly different acquisition parameters.

\begin{figure}[!ht]
  \centering
  \includegraphics[width=0.48\textwidth]{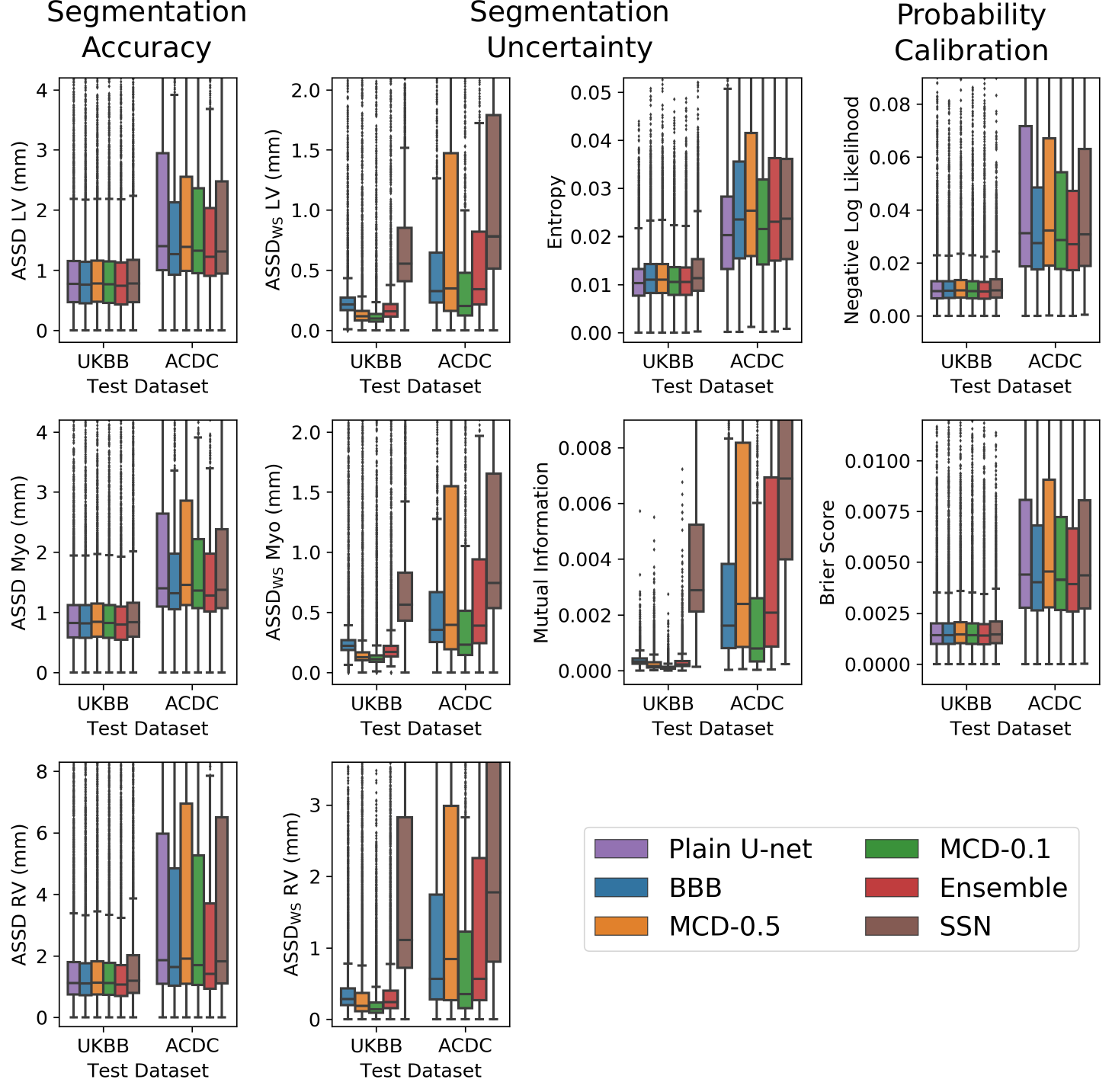}
  \caption{Segmentation accuracy, predictive uncertainty, and probability
  calibration of models trained on the UKBB training dataset and tested on the
  UKBB and ACDC datasets.}
  \label{fig:metrics_correlations}
\end{figure}

As shown in Figure \ref{fig:metrics_correlations}, we observed decreased
segmentation accuracy and increased predictive uncertainty compared to the UKBB
test dataset. In terms of segmentation accuracy and probability calibration, the
methods from the best to the worst are: Deep Ensembles, BBB, MCD-0.1,
SSN, Plain U-net, and MCD-0.5, as shown in Figure
\ref{fig:metrics_correlations} and Supplementary Table S4.
While most metrics are statistically different, some metrics
between the following pairs are not statistically different: Deep Ensembles 
vs BBB, MCD-0.1 vs SSN, SSN vs Plain U-net, and Plain U-net vs MCD-0.5. 
Supplementary Table S5 shows detailed 
results of the pairwise significance tests.

\subsection{Correlations between Uncertainty and Segmentation Accuracy}
To demonstrate the potential utility of uncertainty measures, we evaluated the
Spearman rank correlation between uncertainty measures and segmentation
accuracy. We used the rank correlation instead of linear correlation to reduce
the effects of a potential non-linear relationship between the two quantities.

Supplementary Figure S5 shows that the uncertainty measure with
the strongest correlation with ASSD was ASSD$_\textrm{WS}$ (Spearman correlation
between 0.58 and 0.69) in the case of training and testing on the UKBB
dataset (UKBB $\rightarrow$ UKBB). This is not surprising since the
ASSD$_\textrm{WS}$ calculation was similar to ASSD. Similar observations were
obtained for training on the UKBB dataset and testing on the ACDC dataset (UKBB
$\rightarrow$ ACDC) although the correlations were slightly lower (Spearman
correlation between 0.44 and 0.66). These correlations suggest that the
ASSD$_\textrm{WS}$ uncertainty measure is useful in predicting
the segmentation quality when manual segmentation is not available. 
Other uncertainty measures such as MI or 
Dice$_\textrm{WS}$ may correlate better with other segmentation quality 
metrics such as pixelwise accuracy or Dice (not explored in this work). 
They could also potentially be used as inputs to segmentation algorithms to 
improve segmentation performance.

\begin{figure}[!ht]
   \centering
   \includegraphics[width=0.45\textwidth]{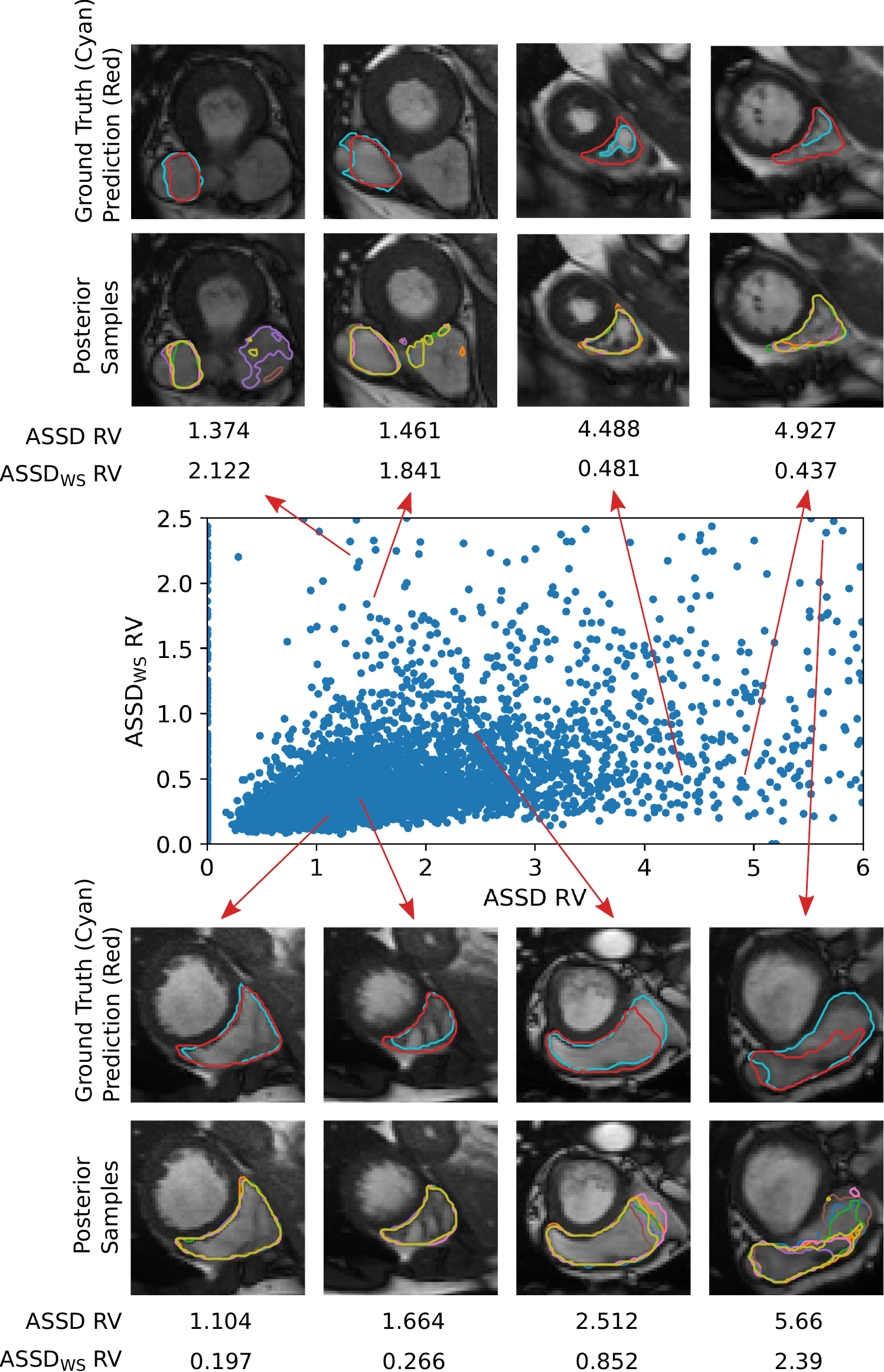}
   \caption{A scatter plot showing the relationship between ASSD and
   ASSD$_\textrm{WS}$ using BBB. Images with manual and predicted segmentations,
   and posterior segmentation samples illustrate varying ASSD and
   ASSD$_\textrm{WS}$. Reproduced by kind permission of UK Biobank \copyright.}
   \vspace{-0.5cm}
   \label{fig:uncertainty_measures}
\end{figure}

Figure \ref{fig:uncertainty_measures} shows representative segmentation results,
posterior prediction samples, and the structural uncertainty measures for RV.
The images with poor segmentation (toward the right side) had greater uncertainty as
measured by Dice$_\textrm{WS}$ and ASSD$_\textrm{WS}$. Note that posterior
prediction samples are not related to inter- or intra-observer variability
but rather show what the network has learned from given data.

\subsection{Uncertainty for Segmentation Quality Control}
In this section, we explored the use of predictive uncertainty estimates to
flag potentially problematic segmentations that require manual review. We view
this task as a classification problem and we aim to use uncertainty measures
to classify segmentations as either good or poor.

While the common segmentation accuracy metrics (Dice, ASSD, 
HD) may not always correlate with true segmentation quality \cite{bernard2018deep}, 
there are no good alternatives to quantify segmentation quality.
Having experts to manually determine whether an automated segmentation is good
or not for a large dataset is time consuming and adds observer noise. Instead,
we used thresholds on segmentation accuracies to achieve this. Based on our
experience and discussions with our clinical collaborators, we believe that
contour or surface distance is more indicative of inaccurate segmentations. As
such, for each method, the predicted segmentation was considered as poor when
the ASSD between the prediction and manual segmentation is greater than the ASSD
between manual observers. We used inter-observer ASSD of 1.17 mm for LV, 1.19 mm
for Myo, and 1.88 mm for RV, based on a recent relatively large-scale study
\cite{bai2018automated}. We then evaluated how well the ASSD$_\textrm{WS}$
uncertainty measure could identify potentially poor segmentation.

\begin{figure*}[!ht]
  \centering
  \includegraphics[width=0.9\textwidth]{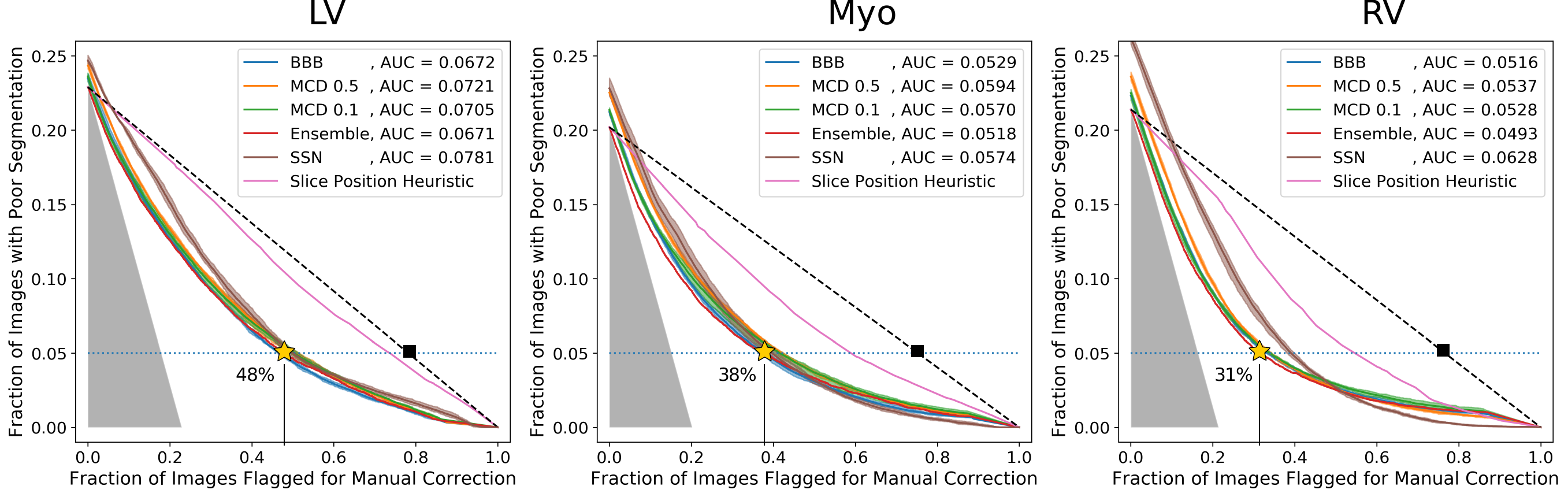}
  \caption{Fraction of images with poor segmentation remaining (based on a
  threshold of ASSD between predicted and ground truth segmentation) after
  flagging images for manual correction using ASSD$_\textrm{WS}$ for BBB, MC
  Dropout, Deep Ensembles, and SSN. Dotted black lines indicate flagging images
  for manual correction randomly. Shaded area shows the ideal region where all images
  flagged for manual correction directly reduce the number of poor
  segmentations.}
  \label{fig:precision_plots_assd}
\end{figure*}

To utilize the ASSD$_\textrm{WS}$ uncertainty measure, a
threshold can be set such that any segmentation with uncertainty above the
threshold is flagged for manual review. This would hopefully result in a
decreased number of poor segmentation in the dataset.
Figure~\ref{fig:precision_plots_assd} shows the fraction of images with poor
segmentation remaining in the dataset and the fraction of images flagged for
manual correction when the uncertainty thresholds were varied, i.e.,
(positives - true positives) \emph{vs} (true positives + false positives), where
positive represents poor segmentation. As we decreased the uncertainty threshold
(top left to bottom right in Figure~\ref{fig:precision_plots_assd}), we flagged
more images for manual correction and the number of images with poor
segmentation was decreased. The first point on the curves corresponds to a
threshold where none of the images are reviewed while the last point
indicates that all the images are reviewed.
This is similar to a receiver operating curve with the consideration that the
total number of positives or images with poor predicted segmentation is
different for each method. This allows for comparison between all the
uncertainty estimation methods. In particular, a curve that is closer to the
bottom left corner or has smaller area under the curve ($\mathrm{AUC}$)
indicates that the algorithm provides better initial segmentation and/or its
uncertainty is a good indicator of segmentation accuracy. Figure
\ref{fig:precision_plots_assd} shows that all the methods performed similarly
for detecting poor LV, Myo, and RV segmentation with Deep Ensembles having
slightly lower $\mathrm{AUC}$ than the others.

The thresholds of the uncertainty measures for flagging images for manual review
can be adjusted depending on the application. This approach provides a way to
identify images to review and may result in substantial time savings. For
example, using the ASSD$_\textrm{WS}$ uncertainty measure for the Deep Ensembles
method, 48\%, 38\%, and 31\% of the images required manual review in order to
reduce the number of images with poor LV, Myo, or RV segmentation to 5\% of the
test dataset, respectively (Figure \ref{fig:precision_plots_assd}). In contrast,
without using uncertainty measures and assuming no other information about the
images is used, approximately 75\% of the images need to be reviewed to achieve
this goal. Furthermore, using the ASSD$_\textrm{WS}$ uncertainty measure
resulted in more time savings compared to a naive approach based on slice
position. Specifically, since the segmentation is usually worse at the base
and/or apex, a naive approach for segmentation quality control is to first
review all the most basal and apical slices, followed by the second-most
basal and apical slices, and so on. We refer to this approach as using the slice
position as a heuristic for segmentation quality control. As shown in Figure
\ref{fig:precision_plots_assd}, using uncertainty measures is more advantageous
than this naive approach as evidenced by a lower $\mathrm{AUC}$ 
and a smaller number of images to review to have 5\% poor segmentation 
remaining. For example, for Myo, AUC for ASSD$_\textrm{WS}$ was
0.052 for Deep Ensembles and 0.080 for the slice position heuristic
method; 38\% of the total images require review when using the ASSD$_\textrm{WS}$ uncertainty
measure compared to 59\% when using the slice position heuristic. For a dataset
of 10,000 subjects each with 10 slices and manual segmentation of 30 seconds
per structure per slice \cite{bai2018automated}, using the ASSD$_\textrm{WS}$
uncertainty measure results in $\sim$940 hours of time savings compared to
reviewing the images randomly, and $\sim$580 hours of time savings compared to
using the slice position heuristic.

\section{Discussion}

\subsection{BBB \emph{vs} MC Dropout \emph{vs} Deep Ensembles \emph{vs} SSN}
In this work, we evaluated and compared different Bayesian and non-Bayesian
methods for estimating uncertainty in neural networks for cardiac MRI
segmentation. Here, we discuss the similarities and differences about how
uncertainty is learned in these methods, what was learned after training, and
relate these differences to the quality of the predictive uncertainties on
out-of-distribution images.

\begin{figure}[ht]
  \centering
  \includegraphics[width=0.45\textwidth]{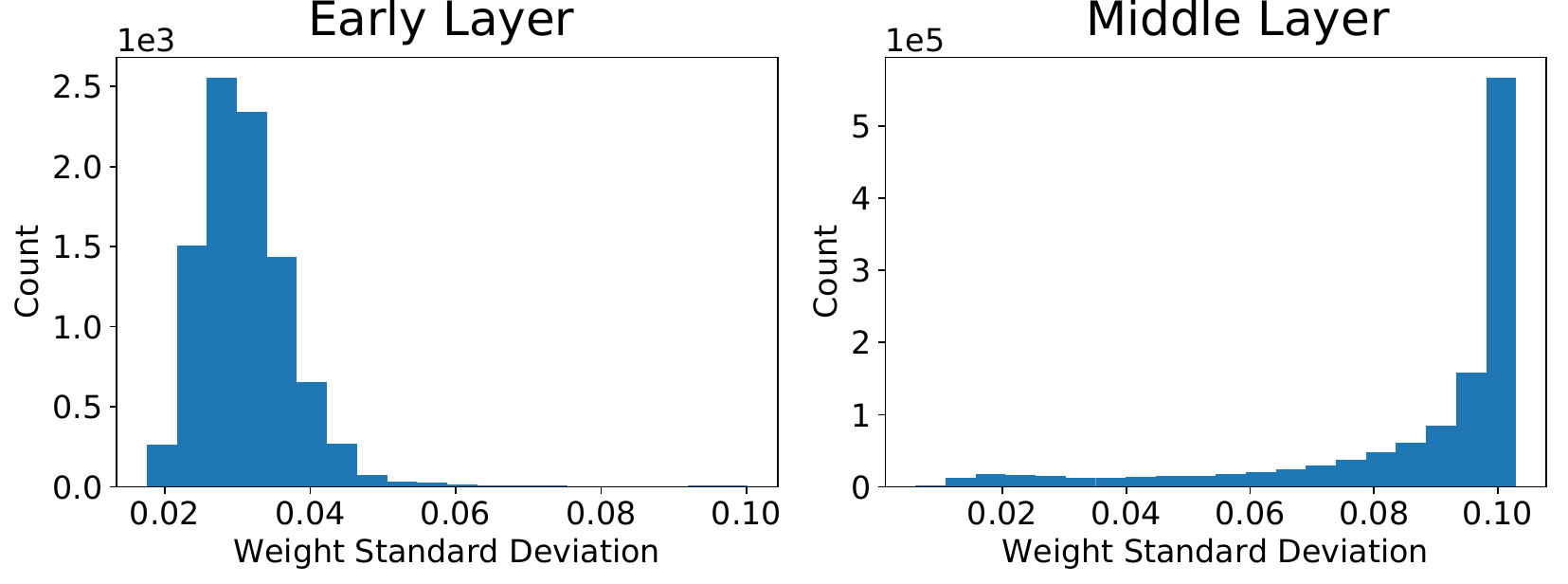}
  \caption{Histogram of the standard deviations of weights in the
  early and middle layers learned by BBB.}
  \label{fig:weight_stats}
\end{figure}

While uncertainty in neural network parameters is learned automatically in BBB,
this can be tuned by changing the dropout rate in MCD. For the UKBB
dataset, a small dropout rate of 0.1 in the middle layers performed better than
the other MCD models in terms of segmentation accuracy and probability
calibration. This is different from other studies which commonly used a dropout
rate of 0.5 \cite{wang2019aleatoric,kendall2017bayesian} and may be because of
the large amount of relatively uniform training data (i.e., images acquired
following the same MR protocol in mainly healthy volunteers and labelled
following the same guidelines).
The dropout rate hyperparameters for MCD obtained through grid search
correspond to the weight uncertainties learned by BBB to some extent. Figure
\ref{fig:weight_stats} shows that the standard deviation of the weights learned
by BBB is lower in the early layers and higher in the middle layers of the
U-net. This is similar to MCD with no dropout in the early layers and
with dropout on the middle layers. Having greater dropout rate in middle layers
is common in other studies \cite{kendall2017bayesian}.

The effect of using different dropout rate in MCD is shown in the experiments with dataset shift
(Figure \ref{fig:metrics_correlations}, Supplementary Tables S4 and S6, MCD-0.1 vs MCD-0.5). A dropout rate of 0.1 yielded
higher segmentation accuracy but lower uncertainty whereas a dropout rate of 0.5
resulted in lower segmentation accuracy but higher predictive uncertainty. BBB
was able to mitigate this issue by learning a mean and standard deviation for
each weight, resulting in comparable or higher segmentation accuracy with
moderate predictive uncertainties between MCD-0.1 and MCD-0.5 (Figure
\ref{fig:metrics_correlations}).

In Deep Ensembles, the uncertainty in the weights is not learned or tuned 
but instead stems from the random initialization of the weights and
stochasticity of the training procedure. Each model in the ensemble would 
learn a local minimum, which are combined to form a prediction and uncertainty
estimate. Deep Ensembles outperformed BBB in terms of segmentation accuracy and
probability calibration in most cases. This may be because the approximate
posterior learned by BBB covered only one (or a few) mode(s) of the true
posterior and a small neighbourhood around each mode, as opposed to multiple
local modes in Deep Ensembles. Based on our observations, the histogram of the
weights learned by each member in Deep Ensembles was very similar to each other.
Techniques such as canonical correlation analysis may be used to compare the
learned feature maps between different members of the ensemble to better 
understand model uncertainty \cite{morcos2018insights}.

The segmentation accuracy and uncertainty measures on images with 
increasing noise and blurring distortions provide some insights into the
different algorithms tested. 
BBB outperformed the other methods in cases of noise distortions probably 
because the Gaussian distribution of the weights is complementary to the 
noise applied on the images (i.e., a Gaussian multiplied by an 
approximately-Gaussian distribution results in a Gaussian distribution). SSN outperformed
the other methods in cases of blurring distortions since blurring is localized
to a small region around each pixel and SSN models the distribution between
pixels. Stretching distortions are complicated structural variations 
and in this case, Deep Ensembles performed the best by relying on weights from several local minima to model uncertainty.
While these simple perturbations (noise, blurring, and stretching distortions) 
may not be realistic, they did highlight cases where BBB or SSN or Deep 
Ensembles are better than others. We tested these algorithms on the ACDC 
dataset for a more realistic comparison of the different algorithms.

Bayesian approaches and Deep Ensembles can be used to improve segmentation 
accuracy on slightly shifted datasets compared to the plain U-net (Supplementary Tables S4 and S5 
show that there is a significant difference between the plain U-net and BBB/Deep Ensembles). While the segmentation accuracies
on the ACDC test dataset using the models trained on UKBB dataset are lower
than that obtained by training and testing on the ACDC dataset
\cite{bernard2018deep}, the algorithms employed in this work may be combined
with other techniques that are specifically designed to solve this problem,
e.g., style augmentation or domain adversarial training \cite{ly2019style,
ganin2016domain}.

\subsection{Pixelwise and Structural Uncertainty Measures}

We introduced pixelwise and structural uncertainty measures to quantify the
predictive uncertainty, and demonstrated the utility of these metrics for
segmentation quality control. Both pixelwise and structural uncertainty measures
can be used depending on the application. As the segmentation problem was
formulated as pixel classification, pixelwise uncertainty measures are
straightforward to obtain. These allow users to visualize which pixels and which
areas are potentially problematic (Figure \ref{fig:distortion_entropy_images}).
However, segmentation is often performed at the image-level slice by
slice. Therefore, image-level uncertainty measures for determining problematic
segmentation are also required. Accordingly, we showed that structural
uncertainty measures were correlated with segmentation accuracy such as ASSD.

Other studies such as \cite{camarasa2021quantitative} 
evaluated uncertainty maps by comparing uncertainty and correctness at 
the per-pixel level. In contrast, we evaluated per-image uncertainty 
measures as a predictor of image-level segmentation quality, which 
is more reflective of the real-world scenario.

It is important to note that the predictive uncertainty measures reflect the
neural network uncertainty, which is different from human uncertainty. An
example is the image with heavy noise in the last row in Figure
\ref{fig:distortion_entropy_images}. It is expected that human observers can
manually segment this image with low observer variability; however, since this
image is very different from the training data, the neural networks were not
able to generate a reasonable segmentation and yielded high predictive entropy
and mutual information for the entire cardiac structure.

\subsection{Segmentation Quality Control}

We showed that the uncertainty measures have moderate to good correlations with
segmentation accuracy. This could have been negatively affected by manual
segmentation noise. Framing segmentation quality control as a binary
classification problem instead of evaluating the correlations or predicting the
segmentation accuracy alleviates the issue of noise in manual segmentation. In
this regard, we defined poor segmentation using a threshold on ASSD between the
predicted and manual segmentation. This definition was adopted based on
discussions with our clinical collaborators; however, it can be modified
depending on the application. For example, other segmentation accuracy metrics
such as Hausdorff distance or misclassification area can be used and the
framework for evaluating uncertainty measures developed in this work may be
applied directly.

Other studies of segmentation quality control include directly predicting
segmentation accuracy or comparing the predicted segmentation to a reference
database. Alba et al. \cite{alba2018automatic} trained a random forest
classifier to predict a binary label of correct or incorrect segmentation. The
examples of correct segmentation were generated using the manual delineation
while the incorrect segmentations were obtained by deforming or translating the
manual segmentations. Robinson et al. \cite{robinson2018neural} used a 3D residual
network to directly predict the Dice of a segmentation from an
image-segmentation pair. The network was trained and tested on a dataset created
using a random forest segmentation algorithm. 
Ruijsink et al. \cite{ruijsink2019fully} trained a CNN for detecting images with artefacts 
or incorrect planning and then excluded these from the segmentation 
pipeline. \cite{ruijsink2019fully} and \cite{sander2020automatic} also 
trained classification models to predict whether a segmentation is good or poor.
These classifiers are agnostic to
how the segmentation was generated. However, these approaches depend on
training with expected segmentation failures, which may be challenging to
incorporate during training. In contrast, our approach used uncertainty measures
to detect poor segmentation and is more explainable without these issues.
Instead of using a learning algorithm to determine segmentation quality, we used
model uncertainty which emerges intrinsically during algorithm training. A
slight limitation of our approach is that sampling during the testing phase
required up to 50x more computation time compared to a single prediction but
this may be accelerated through parallelization.

Additionally, some studies estimate the ``similarity'' of the test image 
and/or predicted segmentation with respect to the training data as a proxy 
for segmentation quality. For example, Gonzalez and Mukhopadhyay \cite{gonzalez2021self} used scores
from a self-supervised task to detect out-of-distribution test images 
(which can then be assumed to have poor segmentation results). 
Galati and Zuluaga \cite{galati2021efficient} trained a convolutional autoencoder to 
reconstruct segmentation maps. Then, the predicted segmentation is fed 
into the autoencoder and segmentation quality measures are calculated 
based on the predicted segmentation and reconstructed prediction.
These generative modelling approaches represent a promising line 
of work and are quite different from the discriminative approaches used in this work. 
One advantage of \cite{gonzalez2021self} is that the algorithm does not 
require ground truth segmentation for training. However, in both approaches, the 
prediction of the segmentation itself is decoupled from the prediction of 
segmentation quality.

Finally, another approach for segmentation quality control is using a modified version 
of Reverse Classification Accuracy to predict the accuracy of an 
image-segmentation pair \cite{robinson2019automated}. This approach requires a
reference database with manual segmentation. Each reference image is registered
to the test image and the associated manual segmentations are warped
accordingly to generate potential segmentations of the test image. Segmentation
quality is estimated by comparing the potential segmentations and algorithm
segmentation. A limitation of this approach is that it requires long time to
predict the segmentation quality, mainly due to the registration steps.

\section{Conclusions}

In this work, we compared Bayesian and non-Bayesian methods, namely BBB, MCD, 
Deep Ensembles, and SSN for segmentation accuracy, probability
calibration, and uncertainty estimates in the context of cardiac MRI
segmentation in cases of various distortions. We found that Deep Ensembles
performed better in terms of segmentation accuracy and probability calibration
on in-distribution and out-of-distribution datasets; BBB outperformed the other
methods on images with noise distortions while SSN outperformed the others
on images with blurring distortions. We showed that ASSD$_\textrm{WS}$ uncertainty
measure was strongly correlated with the segmentation accuracy; using
uncertainty measures can result in substantial time savings by reducing the
number of images that needs manual review for segmentation quality control.

\section*{Acknowledgments}
This research has been conducted using the UK Biobank Resource under Application Number 2964. 
We acknowledge the use of the facilities of Compute Canada. This work was funded by Canadian
Institutes of Health Research (CIHR) MOP: \#93531, Ontario Research Fund and GE Healthcare.
FG is supported by a Banting postdoctoral fellowship.

This work was partly funded by the European Union’s Horizon 2020 research and innovation 
programme under grant agreement No 825903 (euCanSHare project). SEP acts as a paid 
consultant to Circle Cardiovascular Imaging Inc., Calgary, Canada and Servier. SEP 
acknowledges support from the National Institute for Health Research (NIHR)
Biomedical Research Centre at Barts, from the SmartHeart EPSRC programme grant 
(EP/P001009/1) and the London Medical Imaging and AI Centre for Value-Based Healthcare. 
SEP acknowledges support from the CAP-AI programme, London’s first AI enabling programme 
focused on stimulating growth in the capital’s AI sector. SEP, SN and SKP acknowledge the 
British Heart Foundation for funding the manual analysis to create a cardiovascular 
magnetic resonance imaging reference standard for the UK Biobank imaging resource in 5000 
CMR scans (PG/14/89/31194). This project was enabled through access to the Medical 
Research Council eMedLab Medical Bioinformatics infrastructure, supported by the Medical 
Research Council (MR/L016311/1).

\bibliographystyle{IEEEtran}
\bibliography{SegmentationQuality}

\end{document}


\title{\emph{Supplementary Material to} ``Estimating Uncertainty in Neural Networks for 
Cardiac MRI Segmentation: A Benchmark Study''}

\author{Matthew Ng, Fumin Guo, Labonny Biswas, Steffen E. Petersen, Stefan K.
Piechnik, Stefan Neubauer, Graham Wright
}

\markboth{IEEE TRANSACTIONS ON BIOMEDICAL ENGINEERING, VOL. XX, NO. XX, 202X}
{Ng \MakeLowercase{\textit{et al.}}: Estimating Uncertainty in Neural Networks for Cardiac MRI Segmentation: 
A Benchmark Study}

\maketitle

\renewcommand{\thefigure}{S\arabic{figure}}
\renewcommand{\thetable}{S\arabic{table}}

\section{Training Details} \label{sec:more_training_details}
For preprocessing, the input images were cropped to 160$\times$160 pixels from the
center of the original images. The cropped images were normalized by subtracting
the mean and dividing by the variance of the entire training dataset. All the
experiments were repeated 5 times (except for Deep Ensembles) and the resulting
metrics were averaged. Data augmentation was performed, including random
rotation (-60 to 60 degrees), translation (-60 to 60 pixels), and scaling (0.7
to 1.3 times). These models were trained using an Adam optimizer for 50 epochs.
The initial learning rate was set to 1e-4, which was decayed to 1e-5 after
30 epochs.

All the experiments were performed on an Nvidia P100 GPU with 12GB of memory.
SSN, MCD and BBB required approximately 20, 35 and 47 hours for training,
respectively. For these methods, generating 50 prediction samples
during testing required 1.5 seconds for each 2D image. The Deep Ensembles method
consisted of 10 copies of the plain U-net and was trained in parallel
separately. Each plain U-net required $\sim$12 hours for training and $\sim$0.1
seconds for prediction of a 2D image. Calculation of the predictive entropy,
mutual information, Dice$_\textrm{WS}$, and ASSD$_\textrm{WS}$ required
approximately 0.05, 0.15, 0.15, and 0.4 seconds per image, respectively.

\begin{figure*}[!htp]
   \centering
   \includegraphics[width=0.98\textwidth]{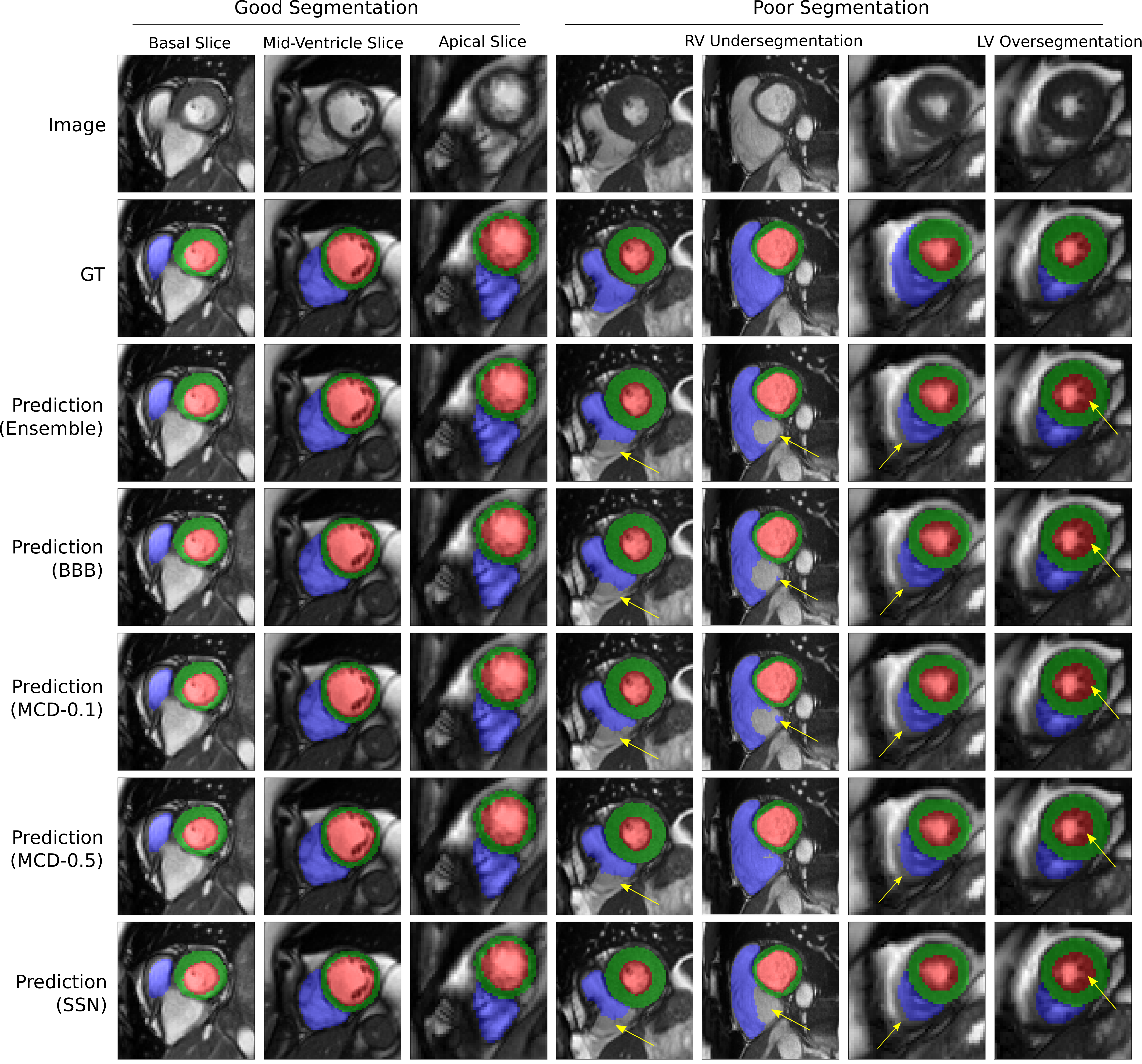}
   \caption{Example segmentation results using the U-net with Deep Ensembles, BBB,  
   MC Dropout (MCD), and SSN. Red = LV, green = Myo, 
   blue = RV. Yellow arrows show area of inaccurate segmentation. Note that all methods 
   have similar results. Reproduced by kind permission of UK Biobank \copyright.}
   \label{fig:sample_results}
 \end{figure*}

 \begin{figure*}[!htp]
   \centering
   \includegraphics[width=0.98\textwidth]{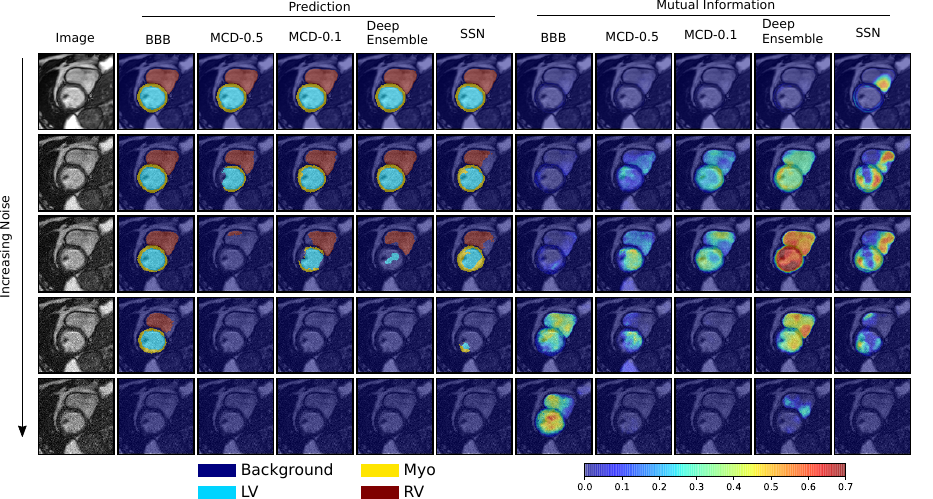}
   \caption{Segmentation predictions and pixelwise uncertainty (mutual information) on images with increasing noise. 
   Segmentation accuracy decreases while predictive uncertainty increases with more noise. 
   Reproduced by kind permission of UK Biobank \copyright.}
   \label{fig:noise_distortion_image}
 \end{figure*}

\begin{figure*}[!htp]
   \centering
   \includegraphics[width=0.98\textwidth]{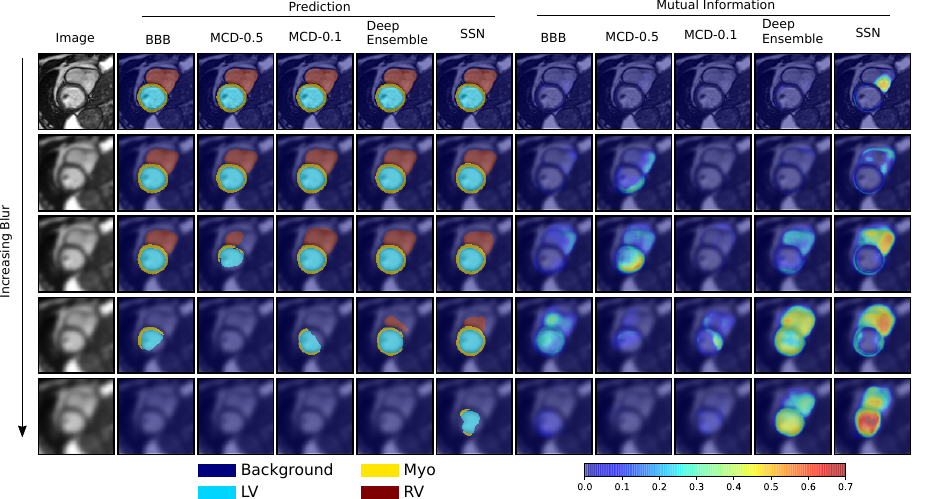}
   \caption{Segmentation predictions and pixelwise uncertainty (mutual information) on images with increasing blur. 
   Segmentation accuracy decreases while predictive uncertainty increases with more blurring. 
   Reproduced by kind permission of UK Biobank \copyright.}
   \label{fig:gaussian_blur_distortion_image}
 \end{figure*}

 \begin{figure*}[!h]
   \centering
   \includegraphics[width=0.98\textwidth]{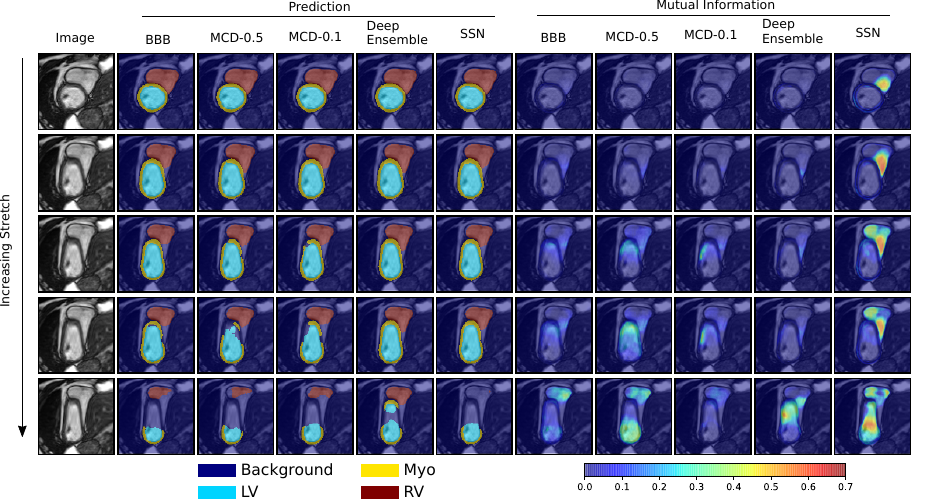}
   \caption{Segmentation predictions and pixelwise uncertainty (mutual information) on images with increasing stretch. 
   Segmentation accuracy decreases while predictive uncertainty increases with
   more stretching. Reproduced by kind permission of UK Biobank \copyright.}
   \label{fig:stretch_distortion_image}
 \end{figure*}

\begin{table*}[h]
  \centering
  \begin{threeparttable}
  \caption{Segmentation accuracy on images with increasing noise, blurring, and 
  stretching distortions. Format: \texorpdfstring{Median\textsubscript{25-75th percentile}}.}
  \label{tab:distortion_accuracy}
  \setlength\tabcolsep{4pt}
  \footnotesize
  \begin{tabularx}{1.0\textwidth}{lllccccc}
  \toprule
               & & 	& 	\multicolumn{5}{c}{Degree of Distortion} \\
               \cmidrule(l{5pt}r{5pt}){4-8}
               & &	&	  0		  &	  1		 &	2		&	3		&	4	\\
               
\midrule
\parbox[t]{4mm}{\multirow{19}{*}{\rotatebox[origin=c]{90}{Noise}}}
& \multirow{6}{*}{Pixelwise Accuracy}
   &    Plain U-net & $.9960_{.9944-.9973}$ &	$.9841_{.9716-.9926}$ &   $.9749_{.9599-.9887}$ &	$.9686_{.9525-.9862}$ &	${.9648_{.9464-.9854}}^*$    \\
   & &	BBB		   & $.9960_{.9944-.9973}$ &	$\mbf{.9929_{.9888-.9956}}$ &	$\mbf{.9905_{.9833-.9947}}$ &	$\mbf{.9862_{.9754-.9932}}$ &	$\mbf{.9735_{.9593-.9883}}$    \\ 			     
   & &	MCD-0.5  	& $.9959_{.9942-.9972}$ &	$.9807_{.9673-.9913}$ &	$.9729_{.9582-.9879}$ &	$.9678_{.9514-.9859}$ &	${.9647_{.9461-.9854}}^*$    \\
   & &	MCD-0.1  	& $.9960_{.9944-.9973}$ &	$.9852_{.9743-.9932}$ &	$.9767_{.9627-.9889}$ &	$.9700_{.9542-.9864}$ &	${.9649_{.9467-.9854}}^*$    \\
   & &  Ensemble    & $\mbf{.9961_{.9945-.9973}}$ &	$.9873_{.9717-.9941}$ &	$.9754_{.9568-.9910}$ &	$.9679_{.9495-.9871}$ &	${.9645_{.9456-.9854}}^*$    \\
   & &  SSN          & $.9960_{.9943-.9972}$ &	$.9811_{.9683-.9916}$ &	$.9725_{.9574-.9878}$ &	$.9675_{.9506-.9858}$ &	${.9648_{.9460-.9854}}^*$    \\
\cmidrule{2-8}
& \multirow{6}{*}{Dice LV}		
   &    Plain U-net & $.963_{.932-.981}$ &	$.653_{.194-.900}$ &   $.224_{.010-.714}$ &	$.031_{.000-.357}$ &	${.000_{.000-.001}}^*$    \\
   &	&  BBB		   & $.964_{.934-.982}$ &	$\mbf{.947_{.863-.976}}$ &	$\mbf{.927_{.701-.972}}$ &	$\mbf{.828_{.307-.961}}$ &	$\mbf{.162_{.000-.753}}$    \\
   & &	MCD-0.5  	& $.963_{.933-.981}$ &	$.657_{.081-.944}$ &	$.235_{.000-.762}$ &	$.000_{.000-.452}$ &	${.000_{.000-.000}}^*$    \\
   & &	MCD-0.1  	& $.963_{.934-.982}$ &	$.714_{.271-.951}$ &	$.356_{.000-.753}$ &	$.024_{.000-.529}$ &	${.000_{.000-.000}}^*$    \\
   & &  Ensemble    & $\mbf{.965_{.936-.983}}$ &	$.873_{.000-.970}$ &	$.000_{.000-.937}$ &	$.000_{.000-.000}$ &	${.000_{.000-.000}}^*$    \\
   & &  SSN          & $.963_{.929-.981}$ &	$.501_{.157-.911}$ &	$.171_{.000-.643}$ &	$.041_{.000-.247}$ &	${.000_{.000-.000}}^*$    \\
\cmidrule{2-8}
& \multirow{6}{*}{ASSD LV}		
   &    Plain U-net & $0.77_{0.47-1.15}$ &	$8.51_{1.79-19.14}$ &   $19.24_{8.03-31.70}$ &	$28.37_{14.20-39.34}$ &	${34.85_{19.52-45.93}}^*$    \\
   &	&  BBB		   & $0.76_{0.45-1.14}$ &	$\mbf{1.07_{0.61-2.05 }}$ &   $\mbf{1.42_{0.69-4.40}}$  &  $\mbf{2.71_{ 0.88-13.65}}$ &	$\mbf{19.44_{4.20-33.37}}$   \\ 			     
   & &	MCD-0.5  	& $0.78_{0.47-1.16}$ &	$8.99_{1.23-21.06}$ &	$18.19_{5.02-31.04}$ &	$27.22_{12.21-37.95}$ &	${34.63_{19.59-45.52}}^*$    \\
   & &	MCD-0.1  	& $0.77_{0.46-1.15}$ &	$8.11_{1.10-15.77}$ &	$15.23_{7.73-28.31}$ &	$25.11_{11.62-37.28}$ &	${34.32_{19.19-45.60}}^*$    \\
   & &  Ensemble    & $\mbf{0.74_{0.43-1.12}}$ &	$2.30_{0.73-22.35}$ &	$20.49_{1.36-38.93}$ &	$32.75_{10.46-45.01}$ &	${35.45_{19.79-	47.01}}^*$    \\ 
   & &  SSN          & $0.78_{0.47-1.17}$ &  $11.09_{1.81-22.62}$ &	$21.95_{7.17-31.65}$ &	$28.74_{15.31-38.76}$ &	${34.49_{19.82-45.70}}^*$    \\
\cmidrule{1-8}

\parbox[t]{4mm}{\multirow{19}{*}{\rotatebox[origin=c]{90}{Gaussian Blur}}}
   & \multirow{6}{*}{Pixelwise Accuracy}
      &    Plain U-net & $.9960_{.9944-.9973}$ &	$.9910_{.9868-.9942}$ & $.9842_{.9759-.9913}$ &	$.9732_{.9609-.9878}$ &	${.9664_{.9494-.9856}}^\dagger$       \\
      & &	BBB		   & $.9960_{.9944-.9973}$ &	$.9922_{.9889-.9948}$ &	$.9864_{.9779-.9921}$ &	$.9726_{.9583-.9876}$ &	${.9656_{.9465-.9855}}^\dagger$       \\
      & &	MCD-0.5  	& $.9959_{.9942-.9972}$ &	$.9894_{.9829-.9936}$ &	$.9779_{.9637-.9904}$ &	$.9676_{.9491-.9869}$ &	${.9647_{.9456-.9854}}^\dagger$       \\
      & &	MCD-0.1  	& $.9960_{.9944-.9973}$ &	$.9915_{.9873-.9945}$ &	$.9847_{.9747-.9917}$ &	$.9713_{.9563-.9875}$ &	${.9652_{.9461-.9854}}^\dagger$       \\
      & &  Ensemble    & $\mbf{.9961_{.9945-.9973}}$ &	$\mbf{.9923_{.9887-.9950}}$ &	$\mbf{.9871_{.9780-.9926}}$ &	$.9730_{.9580-.9881}$ &	${.9655_{.9460-.9855}}^\dagger$       \\
      & &  SSN          & $.9960_{.9943-.9972}$ &	$.9907_{.9857-.9943}$ &	$.9839_{.9743-.9920}$ &	$\mbf{.9748_{.9610-.9893}}$ &	$\mbf{.9682_{.9496-.9867}}$       \\
   \cmidrule{2-8}
   & \multirow{6}{*}{Dice LV}		
      &    Plain U-net & $.963_{.932-.981}$ &	$.910_{.818-.959}$ &   $.833_{.616-.931}$ &	$.535_{.127-.780}$ &	${.093_{.000-.349}}^\dagger$       \\
      &	&  BBB		   & $.964_{.934-.982}$ &	$.936_{.869-.967}$ &	$.883_{.659-.949}$ &	$.466_{.000-.807}$ &	${.000_{.000-.180}}^\dagger$       \\
      & &	MCD-0.5  	& $.963_{.933-.981}$ &	$.922_{.829-.963}$ &	$.726_{.191-.915}$ &	$.066_{.000-.491}$ &	${.000_{.000-.000}}^\dagger$       \\
      & &	MCD-0.1  	& $.964_{.934-.982}$ &	$.928_{.838-.964}$ &	$.864_{.549-.952}$ &	$.356_{.000-.790}$ &	${.000_{.000-.129}}^\dagger$       \\
      & &  Ensemble    & $\mbf{.965_{.936-.983}}$ &	$\mbf{.940_{.876-.970}}$ &	$\mbf{.914_{.775-.964}}$ &	$.688_{.000-.928}$ &	${.000_{.000-.000}}^\dagger$       \\
      & &  SSN          & $.963_{.929-.981}$ &	$.921_{.831-.960}$ &	$.884_{.744-.946}$ &	$\mbf{.726_{.192-.885}}$ &	$\mbf{.146_{.000-.639}}$       \\
   \cmidrule{2-8}
   & \multirow{6}{*}{ASSD LV}		
      &    Plain U-net & $0.77_{0.47-1.15}$ &	$1.50_{0.96-2.86}$ &    $2.91_{1.37-9.46}$  & $12.59_{4.59-23.71}$ &	${27.88_{13.87-39.61}}^\dagger$       \\
      &	&  BBB		   & $0.76_{0.45-1.14}$ &	$1.26_{0.80-2.01}$ &	$2.06_{1.13-5.93}$  & $13.61_{3.03-27.63}$ &	${32.37_{16.40-44.20}}^\dagger$       \\
      & &	MCD-0.5  	& $0.78_{0.47-1.16}$ &	$1.43_{0.88-2.48}$ &	$4.49_{1.48-21.05}$ & $27.73_{9.68-40.34}$ &	${35.11_{19.08-46.65}}^\dagger$       \\
      & &	MCD-0.1  	& $0.76_{0.46-1.15}$ &	$1.35_{0.86-2.31}$ &	$2.22_{1.12-9.11}$  & $14.70_{3.25-31.19}$ &	${33.21_{16.91-45.17}}^\dagger$       \\
      & &  Ensemble    & $\mbf{0.74_{0.43-1.12}}$ &	$\mbf{1.22_{0.74-1.94}}$ &	$\mbf{1.58_{0.89-3.34}}$  & $4.99_{1.47-32.70}$ &	${34.17_{11.37-46.57}}^\dagger$       \\
      & &  SSN          & $0.78_{0.47-1.17}$ &  $1.41_{0.92-2.37}$ &	$1.99_{1.18-3.96}$  & $\mbf{4.49_{1.99-16.66}}$  & $\mbf{21.57_{5.50-37.39}}$        \\
   \cmidrule{1-8}
              
\parbox[t]{4mm}{\multirow{19}{*}{\rotatebox[origin=c]{90}{Stretch}}}
   & \multirow{6}{*}{Pixelwise Accuracy}
       &   Plain U-net & $.9960_{.9944-.9973}$	& $.9921_{.9896-.9940}$ &   $.9879_{.9820-.9925}$ &	$.9841_{.9767-.9912}$ &	${.9729_{.9645-.9852}}^\mathsection$        \\
       & &	BBB		   & $.9960_{.9944-.9973}$	& $.9918_{.9889-.9939}$ &	$.9849_{.9767-.9916}$ &	$.9793_{.9698-.9895}$ &	${.9684_{.9577-.9834}}^\mathsection$        \\
       & &	MCD-0.5  	& $.9959_{.9942-.9972}$	& $.9924_{.9897-.9942}$ &	$.9887_{.9816-.9928}$ &	$.9846_{.9757-.9913}$ &	${.9720_{.9628-.9846}}^\mathsection$        \\
       & &	MCD-0.1  	& $.9960_{.9944-.9973}$	& $.9922_{.9897-.9941}$ &	$.9881_{.9817-.9926}$ &	$.9837_{.9761-.9911}$ &	${.9722_{.9638-.9845}}^\mathsection$        \\
       & & Ensemble    & $\mbf{.9961_{.9945-.9973}}$	& $\mbf{.9925_{.9902-.9944}}$ &	$\mbf{.9894_{.9826-.9933}}$ &	$\mbf{.9857_{.9768-.9923}}$ &	$\mbf{.9730_{.9638-.9864}}$        \\
       & & SSN          & $.9960_{.9943-.9972}$ & $.9919_{.9889-.9940}$ &	$.9860_{.9794-.9920}$ &	$.9813_{.9740-.9902}$ &	${.9706_{.9626-.9834}}^\mathsection$        \\
   \cmidrule{2-8}
   & \multirow{6}{*}{Dice LV}		
       &   Plain U-net & $.963_{.932-.981}$	& $.945_{.895-.969}$ &  $.902_{.762-.963}$ &	$.861_{.697-.955}$ &	${.648_{.402-.838}}^\mathsection$        \\
       & &	BBB		   & $.964_{.934-.982}$	& $.946_{.886-.970}$ &	$.856_{.644-.960}$ &	$.778_{.541-.944}$ &	${.530_{.210-.754}}^\mathsection$        \\
       & &	MCD-0.5  	& $.963_{.933-.981}$	& $.949_{.900-.971}$ &	$.916_{.739-.967}$ &	$.862_{.654-.961}$ &	${.626_{.358-.809}}^\mathsection$        \\
       & &	MCD-0.1  	& $.964_{.934-.982}$	& $.948_{.898-.970}$ &	$.907_{.742-.965}$ &	$.853_{.658-.958}$ &	${.615_{.343-.816}}^\mathsection$        \\
       & & Ensemble    & $\mbf{.965_{.936-.983}}$	& $\mbf{.952_{.910-.972}}$ &	$\mbf{.933_{.799-.969}}$ &	$\mbf{.892_{.722-.967}}$ &	$\mbf{.672_{.442-.859}}$        \\
       & & SSN          & $.963_{.929-.981}$ & $.944_{.876-.969}$ &	$.867_{.670-.961}$ &	$.805_{.598-.946}$ &	${.568_{.295-.780}}^\mathsection$        \\
   \cmidrule{2-8}
   & \multirow{6}{*}{ASSD LV}		
       &   Plain U-net & $0.77_{0.47-1.15}$	& $1.11_{0.79-1.60}$	& $1.58_{0.90-3.53}$	& $2.25_{1.04-5.10}$ &	${6.81_{2.63-14.20}}^\mathsection$        \\
       & &	BBB		   & $0.76_{0.45-1.14}$	& $1.11_{0.78-1.70}$	& $2.20_{0.95-5.68}$	& $3.65_{1.18-8.26}$ &	${9.84_{4.27-21.04}}^\mathsection$        \\
       & &	MCD-0.5  	& $0.78_{0.47-1.16}$	& $1.05_{0.76-1.51}$	& $1.38_{0.81-3.70}$	& $2.10_{0.92-5.72}$ &	${7.50_{2.99-12.36}}^\mathsection$        \\
       & &	MCD-0.1  	& $0.76_{0.46-1.15}$	& $1.08_{0.77-1.56}$	& $1.51_{0.86-3.83}$	& $2.35_{1.00-5.76}$ &	${7.73_{3.15-16.29}}^\mathsection$        \\
       & & Ensemble    & $\mbf{0.74_{0.43-1.12}}$	& $\mbf{1.02_{0.72-1.46}}$	& $\mbf{1.21_{0.75-2.91}}$	& $\mbf{1.66_{0.82-4.62}}$ &	$\mbf{6.36_{1.89-11.22}}$        \\
       & & SSN          & $0.78_{0.47-1.17}$	& $1.11_{0.80-1.72}$	& $2.04_{0.93-5.19}$	& $3.12_{1.15-7.22}$ &	${8.78_{3.96-19.33}}^\mathsection$        \\
\bottomrule
\end{tabularx}
\begin{tablenotes}\footnotesize
   \item[*] statistically different compared to BBB (Wilcoxon signed-rank test, p $<$ 0.05).
   \item[$\dagger$] statistically different compared to SSN (Wilcoxon signed-rank test, p $<$ 0.05).
   \item[$\mathsection$] statistically different compared to Ensemble (Wilcoxon signed-rank test, p $<$ 0.05).
 \end{tablenotes}
\end{threeparttable}
\end{table*}

\begin{table*}[htp]
  \centering
  \begin{threeparttable}
  \caption{Probability calibration on images with increasing noise, blurring, or stretching 
  distortions. Format: \texorpdfstring{Median\textsubscript{25-75th percentile}}. }
  \label{tab:distortion_calibration}
  \setlength\tabcolsep{4pt}
  \footnotesize
  \begin{tabularx}{0.9\textwidth}{lllccccc}
  \toprule
                   & & 	& 	\multicolumn{5}{c}{Degree of Distortion} \\
                   \cmidrule(l{5pt}r{5pt}){4-8}
                   & &	&	  0		  &	  1		 &	2		&	3		&	4	\\
                   
\midrule
\parbox[t]{4mm}{\multirow{12}{*}{\rotatebox[origin=c]{90}{Noise}}}
& \multirow{6}{*}{NLL $(\times 10^{-2})$}
   &    Plain U-net & $0.94_{0.66-1.32}$ &	$8.53_{2.68-19.45}$	& $20.37_{7.63-38.49}$	& $32.64_{13.47-54.87}$	& ${47.85_{19.72-73.98}}^*$	 \\
   & &	BBB		   & $0.95_{0.67-1.31}$ &	$\mbf{1.73_{1.10- 2.70}}	$  & $\mbf{2.35_{1.33- 4.08}}$	& $\mbf{3.54_{ 1.79- 6.58}}$	& $\mbf{8.52_{ 3.81-15.41}}$	 \\ 			     
   & &	MCD-0.5  	& $0.97_{0.69-1.35}$ &	$6.16_{2.70-12.12}$	& $11.13_{4.86-20.18}$	& $17.15_{ 7.76-29.21}$	& ${29.91_{13.40-45.40}}^*$	 \\
   & &	MCD-0.1  	& $0.94_{0.67-1.32}$ &	$4.90_{1.94-11.47}$	& $12.67_{4.72-23.80}$	& $22.00_{ 9.79-37.35}$	& ${38.16_{16.56-59.05}}^*$	 \\
   & &  Ensemble    & $\mbf{0.92_{0.65-1.28}}$ &	$3.54_{1.66- 7.51}	$  & $ 7.40_{2.98-15.80}$	& $13.87_{ 5.29-29.21}$	& ${32.59_{12.42-55.13}}^*$	 \\
   & &  SSN          & $0.97_{0.69-1.39}$ &	$6.71_{2.74-15.81}$  & $16.30_{5.63-32.33}$  & $28.49_{11.37-48.30}$ & ${46.80_{20.20-71.57}}^*$  \\
\cmidrule{2-8}
& \multirow{6}{*}{Brier Score $(\times 10^{-3})$}
   &    Plain U-net & $1.43_{0.99-2.01}$ &	$6.64_{2.92-12.52}$ &   $11.38_{5.03-18.82}$ &	$14.97_{6.60-22.88}$ &	${17.52_{7.27-26.58}}^*$	 \\
   & &  BBB		   & $1.43_{1.00-2.00}$ &	$\mbf{2.55_{1.62- 4.00}}$ &	$\mbf{3.44_{1.94- 5.93}}$ &	$\mbf{5.04_{2.51- 9.03}}$ &	$\mbf{10.45_{4.72-16.48}}$	 \\ 			     
   & &	MCD-0.5  	& $1.47_{1.02-2.06}$ &	$7.53_{3.42-13.17}$ &	$11.28_{5.11-17.95}$ &	$14.38_{6.43-21.71}$ &	${17.13_{7.24-25.99}}^*$	 \\
   & &	MCD-0.1  	& $1.43_{0.99-2.01}$ &	$5.77_{2.57-10.48}$ &	$ 9.91_{4.62-16.20}$ &	$13.46_{6.28-21.03}$ &	${17.14_{7.24-26.03}}^*$	 \\
   & &  Ensemble    & $\mbf{1.40_{0.96-1.96}}$ &	$4.99_{2.30-10.45}$ &	$ 9.57_{3.83-17.19}$ &	$13.90_{5.78-21.92}$ &	${17.37_{7.26-26.34}}^*$	 \\
   & &  SSN          & $1.46_{1.02-2.10}$ &	$7.30_{3.35-12.74}$ &   $11.58_{5.16-18.69}$ &	$14.89_{6.66-22.89}$	&  ${17.38_{7.26-26.61}}^*$  \\
\cmidrule{1-8}

\parbox[t]{4mm}{\multirow{12}{*}{\rotatebox[origin=c]{90}{Gaussian Blur}}}
& \multirow{6}{*}{NLL $(\times 10^{-2})$}
   &    Plain U-net & $0.94_{0.66-1.32}$ &	$2.26_{1.54-3.46}$ &   $4.77_{2.60-8.96}$ &	$12.30_{5.18-22.95}$	& ${24.51_{9.26-41.24}}^\dagger$	 \\
   & &	BBB		   & $0.95_{0.67-1.31}$ &	$\mbf{1.96_{1.41-2.73}}$ &	$3.47_{2.21-5.28}$ &	$ 8.22_{4.38-13.31}$	& ${17.92_{8.53-28.82}}^\dagger$	 \\ 			     
   & &	MCD-0.5  	& $0.97_{0.69-1.35}$ &	$2.71_{1.76-3.99}$ &	$5.59_{2.89-9.69}$ &	$12.59_{5.22-22.19}$	& ${21.42_{8.84-35.15}}^\dagger$	 \\
   & &	MCD-0.1  	& $0.94_{0.67-1.32}$ &	$2.10_{1.47-3.17}$ &	$4.11_{2.39-7.36}$ &	$11.28_{4.96-19.86}$	& ${23.53_{9.88-37.88}}^\dagger$	 \\
   & &  Ensemble    & $\mbf{0.92_{0.65-1.28}}$ &	$1.99_{1.41-2.79}$ &	$\mbf{3.37_{2.17-5.03}}$ &	$\mbf{6.72_{3.87-10.38}}$	& ${13.36_{6.71-21.80}}^\mathparagraph$	 \\
   & &  SSN          & $0.97_{0.70-1.39}$ &  $2.28_{1.53-3.45}$ &	$3.93_{2.14-6.49}$ & $7.12_{3.19-13.22}$	& $\mbf{12.94_{4.85-25.22}}$   \\
\cmidrule{2-8}
& \multirow{6}{*}{Brier Score $(\times 10^{-3})$}
   &    Plain U-net & $1.43_{0.99-2.01}$ &	$3.27_{2.19-4.85}$ &   $5.99_{3.32- 9.63}$ &	$11.43_{5.18-17.25}$ &	${15.81_{6.77-23.88}}^\dagger$	 \\
   & &  BBB		   & $1.43_{1.00-2.00}$ &	$\mbf{2.86_{2.01-4.02}}$ &	$4.99_{3.02- 7.78}$ &	$10.59_{5.08-16.29}$ &	${15.73_{6.93-24.06}}^\dagger$	 \\ 			     
   & &	MCD-0.5  	& $1.47_{1.02-2.06}$ &	$3.92_{2.46-5.98}$ &	$7.98_{3.72-13.57}$ &	$13.96_{5.62-21.96}$ &	${16.85_{7.11-26.23}}^\dagger$	 \\
   & &	MCD-0.1  	& $1.43_{0.99-2.01}$ &	$3.07_{2.08-4.63}$ &	$5.66_{3.17- 9.59}$ &	$11.97_{5.26-18.55}$ &	${16.50_{7.04-25.56}}^\dagger$	 \\
   & &  Ensemble    & $\mbf{1.40_{0.96-1.96}}$ &	$2.88_{1.99-4.12}$ &	$\mbf{4.85_{2.95- 7.55}}$ &	$9.80_{4.85-14.91}$ &	${15.20_{6.65-22.65}}^\dagger$	 \\ 
   & &  SSN          & $1.46_{1.03-2.11}$	&  $3.38_{2.19-5.15}$ &	$5.73_{2.96-9.23}$  &	$\mbf{9.47_{4.00-15.24}}$	&  $\mbf{13.32_{5.39-21.28}}$  \\
\cmidrule{1-8}

\parbox[t]{4mm}{\multirow{12}{*}{\rotatebox[origin=c]{90}{Stretch}}}
& \multirow{6}{*}{NLL $(\times 10^{-2})$}
   &    Plain U-net & $0.94_{0.66-1.32}$ &	$1.96_{1.47-2.66}$ &   $3.34_{1.95-5.73}$ &	$4.88_{2.37-8.67}$ & ${13.35_{6.05-19.84}}^\mathsection$	 \\
   & &	BBB		   & $0.95_{0.67-1.31}$ &	$2.03_{1.53-2.72}$ &	$3.74_{2.19-5.97}$ &	$5.37_{2.82-8.83}$ & ${12.42_{6.07-18.46}}^\mathsection$	 \\ 			     
   & &	MCD-0.5  	& $0.97_{0.69-1.35}$ &	$1.89_{1.42-2.53}$ &	$2.90_{1.90-4.72}$ &	$4.01_{2.35-6.91}$ & $ {9.97_{5.07-14.89}}^\mathsection$	 \\
   & &	MCD-0.1  	& $0.94_{0.67-1.32}$ &	$1.94_{1.45-2.58}$ &	$3.09_{1.92-5.02}$ &	$4.40_{2.38-7.53}$ & ${11.38_{5.57-16.73}}^\mathsection$	 \\
   & &  Ensemble    & $\mbf{0.92_{0.65-1.28}}$ &	$\mbf{1.84_{1.39-2.40}}$ &	$\mbf{2.73_{1.76-4.11}}$ &	$\mbf{3.65_{2.09-5.79}}$ & $\mbf{8.66_{4.32-12.76}}$	 \\
   & &  SSN          & $0.97_{0.69-1.39}$ &	$2.05_{1.52-2.78}$ &	$3.66_{2.10-5.58}$ &	$5.20_{2.65-8.06}$ &	${12.93_{6.76-18.13}}^\mathsection$  \\
\cmidrule{2-8}
& \multirow{6}{*}{Brier Score $(\times 10^{-3})$}
   &    Plain U-net & $1.43_{0.99-2.01}$	& $2.86_{2.17-3.76}$	& $4.48_{2.75-6.92}$	& $6.10_{3.26- 9.35}$ &	${11.58_{6.20-15.56}}^\mathsection$	 \\
   & &  BBB		   & $1.43_{1.00-2.00}$	& $2.96_{2.23-3.97}$	& $5.41_{3.09-8.38}$	& $7.58_{3.90-11.45}$ &	${13.12_{7.03-17.91}}^\mathsection$	 \\ 			     
   & &	MCD-0.5  	& $1.47_{1.02-2.06}$	& $2.77_{2.10-3.69}$	& $4.13_{2.70-6.59}$	& $5.62_{3.29- 9.01}$ &	${11.24_{6.25-15.14}}^\mathsection$	 \\
   & &	MCD-0.1  	& $1.43_{0.99-2.01}$	& $2.82_{2.15-3.71}$	& $4.34_{2.74-6.71}$	& $5.97_{3.31- 9.18}$ &	${11.52_{6.40-15.33}}^\mathsection$	 \\
   & &  Ensemble    & $\mbf{1.40_{0.96-1.96}}$	& $\mbf{2.70_{2.05-3.50}}$	& $\mbf{3.90_{2.52-6.08}}$	& $\mbf{5.26_{2.94- 8.34}}$ &	$\mbf{10.67_{5.49-14.49}}$	 \\ 
   & &  SSN          & $1.46_{1.02-2.10}$	& $2.97_{2.21-4.02}$ & $5.09_{2.97-7.44}$	& $6.89_{3.66-9.73}$  &	${12.11_{6.92-15.61}}^\mathsection$  \\
\bottomrule
\end{tabularx}
\begin{tablenotes}\footnotesize
   \item[*] statistically different compared to BBB (Wilcoxon signed-rank test, p $<$ 0.05).
   \item[$\dagger$] statistically different compared to SSN (Wilcoxon signed-rank test, p $<$ 0.05).
   \item[$\mathparagraph$] not statistically different compared to SSN (Wilcoxon signed-rank test, p $>$ 0.05).
   \item[$\mathsection$] statistically different compared to Ensemble (Wilcoxon signed-rank test, p $<$ 0.05).
 \end{tablenotes}
\end{threeparttable}
\end{table*}

\begin{table*}[htp]
\centering
\begin{threeparttable}
   \caption{Predictive uncertainty measures on images with increasing noise, blurring, or stretching  
   distortions. For each row, the bolded item indicates the degree of 
   distortion with the highest uncertainty.
   Format: \texorpdfstring{Median\textsubscript{25-75th percentile}}.}
   \label{tab:distortion_uncertainty}
   
   \setlength\tabcolsep{4pt}
   \footnotesize
   \begin{tabularx}{0.9\textwidth}{lllccccc}
   \toprule
                    & & 	& 	\multicolumn{5}{c}{Degree of Distortion} \\
                    \cmidrule(l{5pt}r{5pt}){4-8}
                    & &	&	  0		  &	  1		 &	2		&	3		&	4	\\
                    
 \midrule
\parbox[t]{4mm}{\multirow{23}{*}{\rotatebox[origin=c]{90}{Noise}}}
 & \multirow{6}{*}{Pred Entropy $(\times 10^{-2})$}
    &   Plain U-net & $1.03_{0.76-1.32}$ &	$\mbf{1.10_{0.29-1.72}}$ &  $0.47_{0.10-1.32}$ &	$0.13_{0.04-0.60}$ &	$0.02_{0.01-0.05}$     \\
    & &	BBB		   & $1.11_{0.82-1.43}$ &	$1.66_{0.99-2.34}$ &	$1.85_{0.93-2.67}$ &	$\mbf{1.95_{0.72-2.91}}$ &	$1.23_{0.33-2.65}$     \\ 			     
    & &	MCD-0.5  	& $1.10_{0.82-1.43}$ &	$\mbf{2.13_{1.12-3.09}}$ &	$1.65_{0.90-2.77}$ &	$1.11_{0.69-1.96}$ &	$0.51_{0.35-0.75}$     \\
    & &	MCD-0.1  	& $1.06_{0.79-1.36}$ &	$\mbf{1.50_{0.49-2.18}}$ &	$1.00_{0.22-1.95}$ &	$0.40_{0.11-1.31}$ &	$0.08_{0.05-0.22}$     \\
    & & Ensemble    & $1.05_{0.78-1.35}$ &	$\mbf{1.74_{0.42-3.05}}$ &	$0.88_{0.16-2.79}$ &	$0.22_{0.05-1.35}$ &	$0.03_{0.01-0.07}$     \\
    & & SSN          & $1.13_{0.87-1.52}$ &	$\mbf{1.66_{0.71-2.70}}$ &	$0.97_{0.41-2.04}$ &	$0.48_{0.21-1.04}$ &	$0.08_{0.02-0.22}$     \\
 \cmidrule{2-8}
 & \multirow{5}{*}{MI $(\times 10^{-3})$}
    &   BBB		   & $0.32_{0.24-0.43}$ &	$1.26_{0.65-2.46}$   & $2.19_{0.91-4.79}$	 & $3.54_{1.12-7.38}$ &	$\mbf{3.85_{0.75-9.38}}$     \\ 			     
    & &	MCD-0.5  	& $0.16_{0.11-0.29}$ &	$\mbf{4.32_{2.18-6.96}}$   & $4.10_{2.02-7.16}$	 & $2.82_{1.56-5.75}$ &	$1.23_{0.79-2.00}$     \\
    & &	MCD-0.1  	& $0.09_{0.06-0.13}$ &	$\mbf{1.68_{0.44-3.54}}$   & $1.65_{0.26-3.83}$	 & $0.74_{0.14-2.92}$ &	$0.13_{0.06-0.51}$     \\
    & & Ensemble    & $0.22_{0.16-0.34}$ &	$\mbf{5.42_{1.06-14.00}}$	& $3.61_{0.47-14.15}$ & $0.82_{0.14-7.07}$ &	$0.07_{0.03-0.24}$     \\
    & & SSN          & $2.90_{2.11-5.22}$ &	$\mbf{8.95_{3.91-14.29}}$  & $5.49_{2.23-11.45}$ &	$2.63_{1.08-5.99}$ &	$0.38_{0.09-1.17}$     \\
 \cmidrule{2-8}
 & \multirow{5}{*}{Dice\textsubscript{WS} \,\, \textrm{LV}}
    &   BBB		   & $.990_{.984-.994}$	& $.976_{.920-.990}$ &	$.948_{.804-.984}$ &	$.876_{.699-.972}$ &	$\mbf{.820_{.656-.980}}$     \\ 			     
    & &	MCD-0.5  	& $.994_{.990-.997}$	& $.856_{.714-.964}$ &	$\mbf{.835_{.689-.956}}$ &	$.892_{.723-.976}$ &	$.984_{.936-.996}$     \\
    & &	MCD-0.1  	& $.995_{.992-.997}$	& $.931_{.836-.990}$ &	$\mbf{.920_{.813-.996}}$ &	$.968_{.838-1.00}$ &	$1.00_{.988-1.00}$     \\
    & & Ensemble    & $.993_{.987-.996}$	& $\mbf{.862_{.672-.987}}$ &	$.885_{.670-1.00}$ &	$1.00_{.800-1.00}$ &	$1.00_{1.00-1.00}$     \\
    & & SSN          & $.971_{.935-.984}$ & $\mbf{.798_{.678-.902}}$ &	$.842_{.701-.942}$ &	$.920_{.805-.980}$ &	$.990_{.962-.998}$     \\
 \cmidrule{2-8}
 & \multirow{5}{*}{ASSD\textsubscript{WS} \,\, \textrm{LV}}		
    &	BBB		   & $0.21_{0.17-0.27}$	& $0.45_{0.25-1.07}$	& $0.86_{0.32-2.65}$	& $1.90_{0.47-4.67}$ &	$\mbf{2.96_{0.17-6.60}}$     \\	     
    & &	MCD-0.5  	& $0.12_{0.08-0.16}$	& $2.51_{0.62-5.04}$	& $\mbf{2.94_{0.57-5.92}}$	& $1.86_{0.23-5.40}$ &	$0.17_{0.00-1.07}$     \\  
    & & MCD-0.1  	& $0.10_{0.07-0.14}$	& $1.02_{0.17-2.34}$	& $\mbf{1.16_{0.00-2.75}}$	& $0.41_{0.00-2.33}$ &	$0.00_{0.00-0.10}$     \\
    & & Ensemble    & $0.16_{0.11-0.22}$	& $\mbf{3.34_{0.26-7.43}}$	& $2.48_{0.00-8.84}$	& $0.00_{0.00-4.48}$ &	$0.00_{0.00-0.00}$     \\
    & & SSN         & $0.55_{0.41-0.85}$	& $\mbf{4.15_{1.80-7.28}}$	& $3.44_{1.16-6.77}$ & $1.59_{0.55-4.36}$	&  $0.21_{0.01-0.79}$   \\
 \cmidrule{1-8}
 
\parbox[t]{4mm}{\multirow{23}{*}{\rotatebox[origin=c]{90}{Gaussian Blur}}}
 & \multirow{6}{*}{Pred Entropy $(\times 10^{-2})$}
    &   Plain U-net & $1.03_{0.76-1.32}$	& $\mbf{2.01_{1.27-2.58}}$	& $1.98_{0.94-2.82}$	& $1.34_{0.33-2.29}$	& $0.44_{0.16-1.11}$     \\
    & &	BBB		   & $1.11_{0.82-1.43}$	& $2.28_{1.43-3.02}$	& $\mbf{2.47_{1.22-3.69}}$	& $1.71_{0.47-3.07}$	& $0.52_{0.16-1.36}$     \\			     
    & &	MCD-0.5  	& $1.10_{0.82-1.43}$	& $\mbf{2.52_{1.44-3.62}}$	& $2.13_{0.84-3.58}$	& $1.04_{0.33-2.16}$	& $0.33_{0.20-0.65}$     \\
    & &	MCD-0.1  	& $1.06_{0.79-1.37}$	& $2.09_{1.35-2.69}$	& $\mbf{2.11_{0.99-3.12}}$	& $1.29_{0.30-2.40}$	& $0.30_{0.13-0.75}$     \\
    & & Ensemble    & $1.05_{0.78-1.35}$	& $2.31_{1.42-3.11}$	& $\mbf{2.55_{1.17-3.99}}$	& $1.90_{0.44-3.75}$	& $0.67_{0.21-1.87}$     \\
    & & SSN          & $1.13_{0.87-1.53}$ & $2.44_{1.67-3.23}$	& $\mbf{2.50_{1.57-3.52}}$ & $2.14_{1.13-3.28}$ & $1.43_{0.73-2.32}$     \\
 \cmidrule{2-8}
 & \multirow{5}{*}{MI $(\times 10^{-3})$}		
    &	BBB		   & $0.32_{0.24-0.43}$	& $1.64_{0.83-2.90}$	& $3.63_{1.39-6.82}$	 & $\mbf{3.71_{1.07-8.27}}$	& $1.55_{0.36-4.11}$     \\	     
    & &	MCD-0.5  	& $0.16_{0.11-0.29}$	& $3.10_{1.34-5.50}$	& $\mbf{4.03_{1.20-8.87}}$	 & $2.39_{0.57-5.93}$	& $0.64_{0.33-1.64}$     \\
    & &	MCD-0.1  	& $0.09_{0.06-0.13}$	& $0.81_{0.37-1.66}$	& $\mbf{1.43_{0.51-2.89}}$	 & $1.09_{0.26-2.68}$	& $0.28_{0.09-0.81}$     \\
    & & Ensemble    & $0.22_{0.16-0.34}$	& $2.55_{1.09-5.37}$	& $\mbf{5.46_{1.62-11.59}}$ & $5.13_{0.92-14.59}$	& $2.01_{0.41-7.38}$     \\
    & & SSN          & $2.90_{2.10-5.21}$	& $9.74_{6.29-15.59}$& $\mbf{11.89_{7.25-17.71}}$&	$10.90_{5.85-17.76}$ & $7.92_{3.84-12.95}$    \\
 \cmidrule{2-8}
 & \multirow{5}{*}{Dice\textsubscript{WS} \,\, \textrm{LV}}		
    &	BBB		   & $.990_{.984-.994}$	& $.979_{.958-.986}$	& $.944_{.840-.974}$ &	$\mbf{.816_{.688-.988}}$ &	$.964_{0.804-1.00}$     \\
    & &	MCD-0.5  	& $.994_{.990-.997}$	& $.972_{.891-.986}$	& $\mbf{.878_{.708-.980}}$ &	$.911_{.732-1.00}$ &	$1.00_{0.972-1.00}$     \\
    & &	MCD-0.1  	& $.995_{.992-.997}$	& $.988_{.968-.993}$	& $.976_{.898-.990}$ &	$\mbf{.921_{.788-1.00}}$ &	$1.00_{0.956-1.00}$     \\
    & & Ensemble    & $.993_{.987-.996}$	& $.977_{.928-.987}$	& $.945_{.822-.979}$ &	$\mbf{.829_{.610-1.00}}$ &	$.900_{0.700-1.00}$     \\
    & & SSN          & $.971_{.936-.984}$ & $.921_{.768-.952}$	& $.855_{.736-.909}$ &	$\mbf{.765_{.635-.876}}$ &	$.772_{.604-.938}$      \\
 \cmidrule{2-8}
 & \multirow{5}{*}{ASSD\textsubscript{WS} \,\, \textrm{LV}}		
    &	BBB		   & $0.21_{0.17-0.27}$ &	$0.40_{0.31-0.73}$ &	$1.06_{0.44-3.03}$ &	$\mbf{3.40_{0.19-6.34}}$ &	$0.71_{0.00-4.36}$     \\
    & &	MCD-0.5  	& $0.12_{0.08-0.16}$ &	$0.54_{0.26-1.99}$ &	$\mbf{2.35_{0.28-6.10}}$ &	$1.79_{0.00-5.77}$ &	$0.00_{0.00-0.62}$     \\
    & & MCD-0.1  	& $0.10_{0.07-0.14}$ &	$0.22_{0.16-0.47}$ &	$0.43_{0.19-1.45}$ &	$\mbf{1.10_{0.00-2.99}}$ &	$0.00_{0.00-0.62}$     \\
    & & Ensemble    & $0.16_{0.11-0.22}$ &	$0.43_{0.30-0.90}$ &	$0.92_{0.40-4.67}$ &	$\mbf{3.79_{0.00-9.50}}$ &	$0.78_{0.00-6.59}$     \\
    & & SSN          & $0.56_{0.41-0.86}$	&  $1.25_{0.96-4.28}$ & $2.84_{1.59-6.02}$ &	$\mbf{4.98_{2.39-8.40}}$ &	$4.90_{1.53-8.23}$   \\
\cmidrule{1-8}
\parbox[t]{4mm}{\multirow{23}{*}{\rotatebox[origin=c]{90}{Stretch}}}
 & \multirow{6}{*}{Pred Entropy $(\times 10^{-2})$}
    &   Plain U-net & $1.03_{0.76-1.32}$ &	$1.68_{1.30-2.04}$ &    $1.87_{1.04-2.30}$ &	$\mbf{1.95_{0.95-2.46}}$ &	$1.30_{0.60-2.11}$     \\
    & & BBB		   & $1.11_{0.82-1.43}$ &	$1.98_{1.51-2.44}$ &	$\mbf{2.39_{1.21-3.06}}$ &	$2.38_{1.08-3.26}$ &	$1.46_{0.70-2.39}$     \\			     
    & & MCD-0.5  	& $1.10_{0.82-1.43}$ &	$1.76_{1.40-2.25}$ &	$2.19_{1.30-2.82}$ &	$\mbf{2.39_{1.19-3.16}}$ &	$1.72_{0.77-2.91}$     \\
    & & MCD-0.1  	& $1.06_{0.79-1.37}$ &	$1.70_{1.34-2.10}$ &	$2.01_{1.14-2.55}$ &	$\mbf{2.14_{1.05-2.80}}$ &	$1.50_{0.65-2.49}$     \\
    & & Ensemble    & $1.05_{0.78-1.35}$ &	$1.80_{1.40-2.24}$ &	$2.17_{1.25-2.86}$ &	$\mbf{2.33_{1.15-3.18}}$ &	$1.71_{0.75-2.95}$     \\
    & & SSN          & $1.13_{0.87-1.52}$ &	$2.00_{1.59-2.52}$ &	$2.38_{1.53-2.95}$ &	$\mbf{2.47_{1.45-3.14}}$ &	$1.70_{0.91-2.66}$     \\
 \cmidrule{2-8}
 & \multirow{5}{*}{MI $(\times 10^{-3})$}		
    &	BBB		   & $0.32_{0.24-0.43}$	& $1.03_{0.60-1.81}$	& $2.48_{1.02-4.88}$ &	$\mbf{2.92_{1.10-5.68}}$ &	$2.25_{0.80-4.71}$     \\	     
    & &	MCD-0.5  	& $0.16_{0.11-0.29}$	& $0.87_{0.42-1.76}$	& $2.57_{1.02-5.18}$ &	$3.49_{1.35-6.94}$ &	$\mbf{3.64_{1.14-7.55}}$     \\
    & &	MCD-0.1  	& $0.09_{0.06-0.13}$	& $0.36_{0.19-0.72}$	& $1.10_{0.40-2.21}$ &	$1.44_{0.50-2.85}$ &	$\mbf{1.44_{0.43-3.20}}$     \\
    & & Ensemble    & $0.23_{0.16-0.34}$	& $0.91_{0.50-1.67}$	& $2.50_{1.01-5.22}$ &	$3.26_{1.24-6.78}$ &	$\mbf{3.62_{1.19-7.93}}$     \\
    & & SSN          & $2.89_{2.11-5.23}$ & $6.87_{4.54-10.97}$& $9.93_{6.70-13.08}$&	$\mbf{10.72_{7.12-14.13}}$ & $8.53_{4.84-12.87}$  \\      
 \cmidrule{2-8}
 & \multirow{5}{*}{Dice\textsubscript{WS} \,\, \textrm{LV}}		
    &	BBB		   & $.990_{.984-.994}$	& $.983_{.956-.993}$	& $.937_{.834-.982}$ &	$.922_{.802-.970}$ &	$\mbf{.865_{.696-.950}}$     \\
    & &	MCD-0.5  	& $.994_{.990-.997}$	& $.989_{.965-.996}$	& $.951_{.864-.991}$ &	$.928_{.838-.980}$ &	$\mbf{.870_{.724-.937}}$     \\
    & &	MCD-0.1  	& $.995_{.992-.997}$	& $.992_{.979-.997}$	& $.971_{.919-.995}$ &	$.960_{.895-.989}$ &	$\mbf{.918_{.814-.968}}$     \\
    & & Ensemble    & $.993_{.987-.996}$	& $.987_{.968-.995}$	& $.960_{.900-.993}$ &	$.947_{.877-.990}$ &	$\mbf{.899_{.737-.958}}$     \\
    & & SSN          & $.971_{.936-.984}$	& $.930_{.803-.974}$	& $.828_{.688-.925}$ &	$.808_{.677-.909}$ &	$\mbf{.741_{.619-.851}}$     \\
 \cmidrule{2-8}
 & \multirow{5}{*}{ASSD\textsubscript{WS} \,\, \textrm{LV}}		
    &	BBB		   & $0.21_{0.17-0.27}$	& $0.33_{0.19-0.63}$ &	$0.93_{0.34-2.12}$ &	$1.18_{0.47-2.53}$ &	$\mbf{1.80_{0.68-3.36}}$     \\
    & &	MCD-0.5  	& $0.12_{0.08-0.16}$	& $0.20_{0.10-0.50}$ &	$0.74_{0.18-1.81}$ &	$1.12_{0.33-2.21}$ &	$\mbf{2.00_{0.90-3.34}}$     \\
    & & MCD-0.1  	& $0.10_{0.07-0.14}$	& $0.16_{0.09-0.31}$ &	$0.45_{0.13-1.03}$ &	$0.64_{0.19-1.30}$ &	$\mbf{1.11_{0.47-1.95}}$     \\
    & & Ensemble    & $0.16_{0.11-0.22}$	& $0.28_{0.16-0.48}$ &	$0.62_{0.20-1.37}$ &	$0.81_{0.25-1.84}$ &	$\mbf{1.58_{0.56-3.92}}$     \\
    & & SSN          &$0.56_{0.41-0.85}$  & $1.09_{0.55-3.31}$	&  $2.93_{1.16-6.01}$ &	$3.28_{1.47-6.33}$ &	$\mbf{4.45_{2.44-7.02}}$     \\
\bottomrule
\end{tabularx}
\end{threeparttable}
\end{table*}

\begin{table*}[ht]
    \begin{threeparttable}
    \caption{Segmentation accuracy and probability calibration metrics on the ACDC and 
               UKBB datasets using models trained with UKBB dataset. Format: 
             \texorpdfstring{Median\textsubscript{25-75th percentile}}.}
    \vspace{-0.3cm}
    \label{tab:acdc_accuracy}
    \centering
    \setlength\tabcolsep{2pt}
    \scriptsize
    \begin{tabularx}{1.02\textwidth}{lcccccccccccc}
      \toprule
                     & Pixelwise Accuracy $\uparrow$	& \multicolumn{3}{c}{Dice $\uparrow$}& \multicolumn{3}{c}{ASSD (mm) $\downarrow$} & NLL  $\downarrow$  					& 	BS  $\downarrow$	\\
                     \cmidrule(l{5pt}r{5pt}){3-5} \cmidrule(l{5pt}r{5pt}){6-8} 
                     &   &    LV 	    &   Myo   	&    RV     &    LV      &  Myo  	&   RV   & $(\times 10^{-2})$					& 	$(\times 10^{-3})$	  \\     
     \midrule
     \multicolumn{2}{l}{{\bf UKBB $\rightarrow$ UKBB}}  \\
     Plain U-net & $.9960_{.9944-.9973}$	& $.963_{.932-.981}$	& $.899_{.847-.929}$	& $.929_{.850-.964}$	& $0.77_{0.46-1.15}$	& $0.82_{0.58-1.12}$	& $1.12_{0.74-1.79}$ & $0.94_{0.66-1.32}$	& $1.43_{0.99-2.01}$  \\
     BBB	   & $.9960_{.9944-.9973}$	& $.964_{.934-.982}$	& $.899_{.849-.930}$	& $.931_{.853-.966}$	& $0.76_{0.45-1.14}$	& $0.82_{0.57-1.12}$	& $1.10_{0.72-1.76}$ & $0.95_{0.67-1.31}$	& $1.43_{1.00-2.00}$  \\		     
     MCD-0.1   & $.9960_{.9944-.9973}$	& $.964_{.934-.982}$	& $.899_{.847-.930}$	& $.931_{.850-.965}$	& $0.77_{0.46-1.15}$	& $0.82_{0.58-1.12}$	& $1.11_{0.73-1.77}$ & $0.94_{0.67-1.31}$	& $1.43_{0.99-2.01}$  \\
     MCD-0.5   & $.9959_{.9942-.9972}$	& $.963_{.933-.981}$	& $.896_{.845-.927}$	& $.929_{.847-.964}$	& $0.78_{0.47-1.16}$	& $0.84_{0.60-1.15}$	& $1.13_{0.74-1.82}$ & $0.97_{0.69-1.35}$	& $1.47_{1.02-2.06}$  \\
     Ensemble  & $\mbf{.9961_{.9945-.9974}}$	& $\mbf{.965_{.936-.983}}$	& $\mbf{.902_{.852-.933}}$	& $\mbf{.935_{.863-.967}}$	& $\mbf{0.74_{0.43-1.13}}$	& $\mbf{0.80_{0.55-1.10}}$	& $\mbf{1.07_{0.69-1.71}}$ & $\mbf{0.92_{0.65-1.28}}$	& $\mbf{1.40_{0.97-1.96}}$   \\
      SSN        & $.9960_{.9943-.9973}$ & $.963_{.929-.981}$  &$.896_{.839-.928}$     & $.921_{.820-.961}$ & $0.78_{0.47-1.17}$ & $0.84_{0.59-1.16}$ & $1.19_{0.79-2.02}$ & $0.97_{0.69-1.39}$ & $1.46_{1.02-2.10}$  \\
     \midrule
     \multicolumn{2}{l}{{\bf UKBB $\rightarrow$ ACDC}} \\
     Plain U-net   & $.9880_{.9790-.9923}$	& $.927_{.781-.960}$	& $.833_{.728-.871}$	& $.870_{.600-.938}$	& $1.40_{1.00-2.94}$	& $1.40_{1.10-2.64}$	& $1.86_{1.09-5.96}$ & $3.12_{1.87-7.17}$	& $4.39_{2.77-8.05}$   \\
     BBB		   & $.9888_{.9814-.9927}$	& $.938_{.856-.964}$	& $.843_{.769-.876}$	& $.884_{.615-.943}$	& $1.27_{0.93-2.12}$	& $1.32_{1.05-1.97}$	& $1.64_{1.03-4.84}$ & $2.76_{1.76-4.85}$	& $4.03_{2.64-6.81}$   \\ 
     MCD-0.1   & $.9885_{.9803-.9927}$	& $.935_{.828-.962}$	& $.838_{.754-.873}$	& $.879_{.600-.943}$	& $1.33_{0.95-2.36}$	& $1.36_{1.07-2.21}$	& $1.69_{1.05-5.27}$ & $2.87_{1.78-5.42}$	& $4.15_{2.67-7.22}$   \\
     MCD-0.5   & $.9876_{.9751-.9923}$	& $.930_{.808-.960}$	& $.819_{.669-.866}$	& $.862_{.508-.939}$	& $1.39_{0.99-2.55}$	& $1.46_{1.12-2.85}$	& $1.91_{1.09-6.96}$ & $3.22_{1.91-6.71}$	& $4.53_{2.79-9.06}$   \\
     Ensemble  & $\mbf{.9892_{.9822-.9930}}$	& $\mbf{.941_{.859-.965}}$	& $\mbf{.847_{.773-.878}}$	& $\mbf{.907_{.716-.953}}$	& $\mbf{1.23_{0.90-2.03}}$	& $\mbf{1.28_{1.01-1.97}}$	& $\mbf{1.41_{0.93-3.70}}$ & $\mbf{2.71_{1.73-4.72}}$	& $\mbf{3.94_{2.59-6.67}}$    \\
      SSN        & $.9882_{.9788-.9926}$ & $.934_{.807-.962}$ &$.837_{.736-.871}$      & $.860_{.533-.937}$	& $1.31_{0.95-2.48}$ & $1.38_{1.07-2.38}$ & $1.83_{1.11-6.51}$ & $3.08_{1.89-6.30}$ & $4.36_{2.74-8.05}$   \\
     \bottomrule
    \end{tabularx}
  \end{threeparttable}
  \end{table*}

  \begin{table*}[htp]
   \centering  
   \begin{threeparttable}
   \caption{p-values from the pairwise Wilcoxon signed-rank test comparing
   segmentation accuracy and calibration of different models trained on the 
   UKBB dataset and tested on the ACDC dataset (NS = Not significant, p $>$ 0.05).}
   \vspace{-0.3cm}
   \label{tab:acdc_accuracy_significance}
   \setlength\tabcolsep{3pt}
   \footnotesize
   \begin{tabularx}{0.88\textwidth}{lllcccccccccccc}
     \toprule
                    & & & Pixelwise Accuracy $\uparrow$	& \multicolumn{3}{c}{Dice $\uparrow$}& \multicolumn{3}{c}{ASSD (mm) $\downarrow$} & NLL  $\downarrow$  					& 	BS  $\downarrow$	\\
                    \cmidrule(l{5pt}r{5pt}){5-7} \cmidrule(l{5pt}r{5pt}){8-10} 
                    & & &  &    LV 	    &   Myo   	&    RV     &    LV      &  Myo  	&   RV   & $(\times 10^{-2})$					& 	$(\times 10^{-3})$	  \\     
    \midrule
    \multicolumn{3}{l}{{\bf UKBB $\rightarrow$ ACDC}} \\
    Plain U-net &vs &BBB      & $<$0.0001 & $<$0.0001 & $<$0.0001 & $<$0.0001 & $<$0.0001 & $<$0.0001 & $<$0.0001 & $<$0.0001 & $<$0.0001 \\
    Plain U-net &vs &MCD-0.1  & $<$0.0001 & $<$0.0001 & $<$0.0001 & 0.0027    & $<$0.0001 & $<$0.0001 & $<$0.0001 & $<$0.0001 & $<$0.0001 \\
    Plain U-net &vs &MCD-0.5  & $<$0.0001 & ns        & $<$0.0001 & $<$0.0001 & $<$0.0001 & $<$0.0001 & $<$0.0001 & ns        & $<$0.0001 \\
    Plain U-net &vs &Ensemble & $<$0.0001 & $<$0.0001 & $<$0.0001 & $<$0.0001 & $<$0.0001 & $<$0.0001 & $<$0.0001 & $<$0.0001 & $<$0.0001 \\
    Plain U-net &vs &SSN      & ns        & $<$0.0001 & 0.0006    & 0.0006    & $<$0.0001 & $<$0.0001 & ns        & 0.0002    & ns        \\
    BBB		    &vs &MCD-0.1  & $<$0.0001 & $<$0.0001 & $<$0.0001 & 0.0003    & $<$0.0001 & $<$0.0001 & 0.0002    & $<$0.0001 & $<$0.0001 \\
    BBB		    &vs &MCD-0.5  & $<$0.0001 & $<$0.0001 & $<$0.0001 & $<$0.0001 & $<$0.0001 & $<$0.0001 & $<$0.0001 & $<$0.0001 & $<$0.0001 \\
    BBB		    &vs &Ensemble & $<$0.0001 & ns        & ns        & $<$0.0001 & ns        & ns        & $<$0.0001 & $<$0.0001 & $<$0.0001 \\
    BBB		    &vs &SSN      & $<$0.0001 & 0.0074    & $<$0.0001 & $<$0.0001 & 0.0037    & $<$0.0001 & $<$0.0001 & $<$0.0001 & $<$0.0001 \\
    MCD-0.1     &vs &MCD-0.5  & $<$0.0001 & $<$0.0001 & $<$0.0001 & $<$0.0001 & $<$0.0001 & $<$0.0001 & $<$0.0001 & $<$0.0001 & $<$0.0001 \\
    MCD-0.1     &vs &Ensemble & $<$0.0001 & $<$0.0001 & $<$0.0001 & $<$0.0001 & $<$0.0001 & $<$0.0001 & $<$0.0001 & $<$0.0001 & $<$0.0001 \\
    MCD-0.1     &vs &SSN      & $<$0.0001 & ns        & ns        & $<$0.0001 & ns        & ns        & $<$0.0001 & $<$0.0001 & $<$0.0001 \\
    MCD-0.5     &vs &Ensemble & $<$0.0001 & $<$0.0001 & $<$0.0001 & $<$0.0001 & $<$0.0001 & $<$0.0001 & $<$0.0001 & $<$0.0001 & $<$0.0001 \\
    MCD-0.5     &vs &SSN      & $<$0.0001 & $<$0.0001 & $<$0.0001 & 0.0009    & $<$0.0001 & $<$0.0001 & $<$0.0001 & ns        & $<$0.0001 \\
    Ensemble    &vs &SSN      & $<$0.0001 & ns        & $<$0.0001 & $<$0.0001 & ns        & $<$0.0001 & $<$0.0001 & $<$0.0001 & $<$0.0001 \\
    \bottomrule
   \end{tabularx}
 \end{threeparttable}
 \end{table*}

\begin{table*}[ht]
  \centering
  \begin{threeparttable}
    \caption{Predictive uncertainty measures on the ACDC and UKBB datasets using model 
            trained with UKBB dataset. Format: 
            \texorpdfstring{Median\textsubscript{25-75th percentile}}.}
    \vspace{-0.3cm}
    \label{tab:acdc_uncertainty}
    \setlength\tabcolsep{4pt}
    \scriptsize
    \begin{tabularx}{0.97\textwidth}{lcccccccc}
    \toprule
                    & Pred Entropy 					& 	MI   & \multicolumn{3}{c}{Dice\textsubscript{WS}}  & \multicolumn{3}{c}{ASSD\textsubscript{WS}}  \\
                    \cmidrule(l{5pt}r{5pt}){4-6}	\cmidrule(l{5pt}r{5pt}){7-9}
                    & $(\times 10^{-2})$					& 	$(\times 10^{-3})$	  &    LV 	    &   Myo   	&    RV     &    LV      &  Myo  	&   RV   		\\
    \midrule
      \multicolumn{2}{l}{{\bf UKBB $\rightarrow$ UKBB}} \\
      Plain U-net     & $1.03_{0.77-1.33}$ & N/A  & N/A  & N/A  & N/A  & N/A  & N/A  & N/A \\
      BBB       & $1.11_{0.83-1.43}$	& $0.32_{0.24-0.43}$ & $.990_{.984-.994}$	& $.974_{.964-.979}$	& $.982_{.964-.991}$ & $0.214_{0.167-0.273}$ & $0.222_{0.187-0.270}$ & $0.286_{0.200-0.432}$     \\
      MCD-0.1   & $1.06_{0.79-1.37}$	& $0.09_{0.06-0.13}$ & $.995_{.992-.997}$	& $.987_{.982-.989}$	& $.991_{.981-.996}$ & $0.099_{0.071-0.137}$ & $0.111_{0.088-0.142}$ & $0.142_{0.091-0.238}$     \\
      MCD-0.5   & $1.10_{0.82-1.43}$	& $0.16_{0.11-0.29}$ & $.994_{.990-.997}$	& $.985_{.979-.988}$	& $.988_{.970-.994}$ & $0.116_{0.083-0.162}$ & $0.127_{0.099-0.166}$ & $0.188_{0.117-0.372}$     \\
      Ensemble  & $1.06_{0.79-1.36}$	& $0.22_{0.16-0.34}$ & $.993_{.988-.996}$	& $.980_{.971-.985}$	& $.985_{.967-.993}$ & $0.159_{0.113-0.219}$ & $0.170_{0.134-0.222}$ & $0.245_{0.155-0.404}$     \\
        SSN       & $1.14_{0.87-1.53}$ & $2.89_{2.11-5.24}$ & $.971_{.936-.984}$	& $.935_{.895-.947}$	& $.914_{.794-.959}$ & $0.554_{0.408-0.851}$ & $0.564_{0.430-0.827}$ & $1.113_{0.725-2.826}$     \\
      \midrule
      \multicolumn{2}{l}{{\bf UKBB $\rightarrow$ ACDC}} \\
      Plain U-net     & $2.02_{1.32-2.82}$ & N/A  & N/A  & N/A  & N/A  & N/A  & N/A  & N/A \\
      BBB		& $2.35_{1.55-3.55}$	& $1.61_{0.80-3.83}$ & $.985_{.964-.991}$	& $.961_{.923-.971}$	& $.960_{.840-.984}$ & $0.327_{0.230-0.646}$ & $0.353_{0.254-0.668}$ & $0.567_{0.281-1.741}$     \\ 
      MCD-0.1   & $2.15_{1.42-3.18}$	& $0.79_{0.32-2.60}$ & $.991_{.974-.995}$	& $.975_{.944-.983}$	& $.974_{.876-.991}$ & $0.201_{0.124-0.478}$ & $0.231_{0.146-0.514}$ & $0.356_{0.159-1.228}$     \\
      MCD-0.5   & $2.53_{1.60-4.15}$	& $2.40_{0.84-8.19}$ & $.983_{.925-.993}$	& $.956_{.868-.977}$	& $.940_{.755-.984}$ & $0.350_{0.161-1.473}$ & $0.396_{0.192-1.549}$ & $0.845_{0.269-2.986}$     \\
      Ensemble  & $2.30_{1.50-3.63}$	& $2.08_{0.86-6.94}$ & $.985_{.948-.992}$	& $.957_{.900-.971}$	& $.958_{.814-.986}$ & $0.343_{0.214-0.819}$ & $0.390_{0.242-0.938}$ & $0.570_{0.268-2.257}$     \\                   
        SSN       & $2.36_{1.53-3.61}$ & $6.89_{3.98-14.30}$ &$.961_{.878-.981}$ & $.917_{.830-.938}$	& $.876_{.731-.947}$ & $0.780_{0.513-1.791}$ & $0.744_{0.536-1.654}$	& $1.774_{0.811-3.671}$     \\
    \bottomrule
    \end{tabularx}
  \end{threeparttable}
  \end{table*}
  
  \begin{figure*}[!h]
   \centering
   \includegraphics[width=0.68\textwidth]{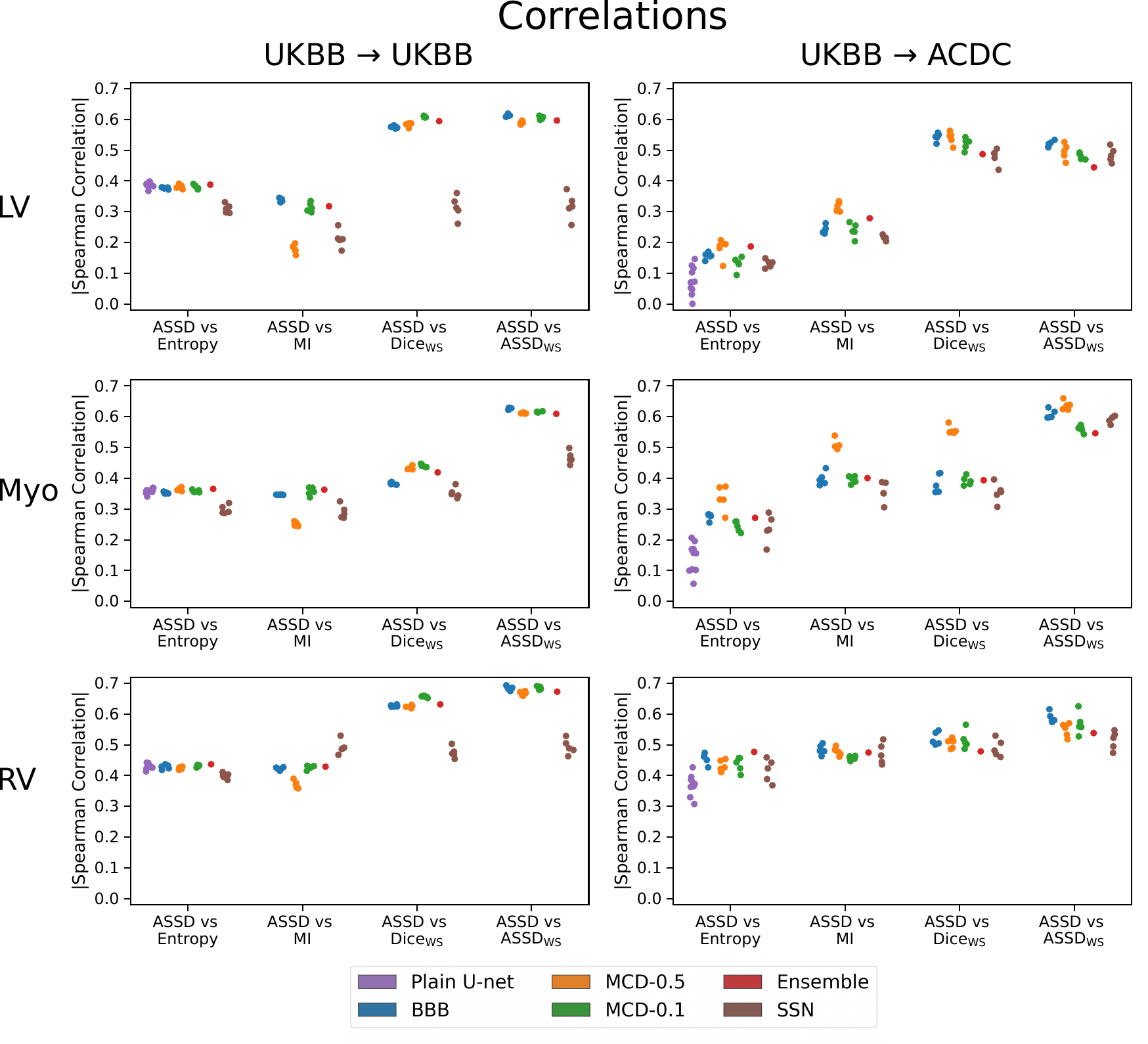}
   \caption{Spearman correlations between segmentation accuracy (ASSD) and 
   uncertainty measures (entropy, mutual information, Dice$_\textrm{WS}$, 
   ASSD$_\textrm{WS}$) on the UKBB and ACDC datasets. Each point represents one run.}
   \label{fig:correlations}
   \end{figure*}